\documentstyle [epsf,12pt]{article}

\catcode`\@=11
\def\makepreprititle{\par
  \begingroup
  \def\thefootnote{\fnsymbol{footnote}}
  \def\
@makefnmark{\hbox 
  to 0pt{$^{\@thefnmark}$\hss}} 
  \if@twocolumn 
  \twocolumn[\@makepreprititle] 
  \else \newpage
  \global\@topnum\z@ 
  \@makepreprititle \fi\thispagestyle{empty}\@thanks
  \endgroup
  \setcounter{footnote}{0}
  \let\makepreprititle\relax
  \let\@makepreprititle\relax
  \gdef\@thanks{}\gdef\@author{}\gdef\@title{}
  \gdef\@preprintnumber{}\gdef\@preprintdate{}\gdef\subtitle{}
  \let\thanks\relax}
\def\preprintnumber#1{\gdef\@preprintnumber{#1}}
\def\preprintdate#1{\gdef\@preprintdate{#1}}
\def\subtitle#1{\gdef\@subtitle{#1}}
\def\@makepreprititle{\newpage
{\def\baselinestretch{1}
  \begin{flushright} \@preprintnumber \par
  \@preprintdate \end{flushright} } \par
  \begin{center}
\vskip 1.5em
  {\LARGE \@title \par} \vskip 2.5em 
  {\Large \lineskip .5em
  \begin{tabular}[t]{c}\@author 
  \end{tabular}\par}
  \vskip 1em {\large \@date} \end{center}
  \par
  \vfil} 
\date{\sl Department of Physics, Tohoku University\\Sendai, 980 Japan}
\preprintdate{}
\preprintnumber{}
\subtitle{}
\def\abstract{\if@twocolumn
\section*{Abstract}
\else \normalsize 
\begin{center}
{\bf Abstract\vspace{-.5em}\vspace{0pt}} 
\end{center}
\quotation 
\addtocounter{page}{-1}
\fi}
\def\endabstract{\if@twocolumn\else\endquotation\fi}
\catcode`\@=12
\def\thebibliography#1{\section*
 {References					% PLB/NPB style
 \markboth{REFERENCES}{REFERENCES}}\list
 {[\arabic{enumi}]}				% PLB/NPB style
 {\settowidth\labelwidth{[#1]}\leftmargin\labelwidth
 \advance\leftmargin\labelsep
 \usecounter{enumi}}
 \def\newblock{\hskip .11em plus .33em minus -.07em}
 \sloppy \sfcode`\.=1000\relax}
 
\renewcommand{\theequation}{\thesection.\arabic{equation}}
\newcommand{\cleqn}{\setcounter{equation}{0}}
\newcommand{\clfn}{\setcounter{footnote}{0}}

\renewcommand{\thefootnote}{\fnsymbol{footnote}}

\begin{document}

%%%%%   Cover Page

\preprintnumber{TU--489}
\preprintdate{August, 1995}

\vspace{10mm}

\title{
Bern-Kosower Rule for Scalar QED
}

\author{K.~Daikouji, M.~Shino, and Y.~Sumino}
\date{\sl  Department of Physics, Tohoku University, Sendai, 980 Japan}
\makepreprititle
%%%%% abstract

\vspace{15mm}

\begin{abstract}
\normalsize

We derive a full Bern-Kosower-type rule for scalar QED starting from
quantum field theory: we derive a set of rules for calculating
$S$-matrix elements for any processes at any order of the coupling
constant. 
Gauge-invariant set of diagrams in general is first written in the 
worldline path-integral expression.
Then we integrate over $x(\tau)$, and the resulting expression is
given in terms of correlation function on the worldline
$\left< x(\tau) x(\tau') \right>$.
Simple rules to decompose the correlation function
into basic elements are obtained.
Gauge transformation known as integration by parts technique can be used 
to reduce the number of independent terms before integration over
proper-time variables.
The surface terms can be omitted provided the external scalars are
on-shell. 
Also, we clarify correspondence to the conventional Feynman rule, which
enabled us to avoid any ambiguity coming from the infinite
dimensionality of the path-integral approach.

\end{abstract} 
\vspace{1cm}

\vfil

%%%%% text
\newpage
\baselineskip 22pt

\section{Introduction}
\cleqn

Recently, Bern and Kosower derived from superstring theory
a powerful method for calculating
one-loop $S$-matirx elements for QCD processes.\cite{bk}
Although the new rule had reduced the amount of work
required in the calculation greatly, it had little resemblance to the
conventional Feynman rule, and to date, the complete Bern-Kosower rule
has not been derived from quantum field theory (QCD).
The equivalence of the Bern-Kosower rule and the conventional Feynman
rule has been shown only in some concrete examples.\cite{bd}
Moreover, practical problems are that since the Bern-Kosower rule has
been derived from the string theory, it is difficult to include massive
particles and also multi-loop generalizations do not readily
lead to simple calculational tools.\cite{roland}

As for the approach from the quantum field theory, there has been
some progress.
Bern-Kosower-type rules for calculating
one-loop effective actions for both abelian and non-abelian gauge
theories have been derived from quantum field theories and studied
extensively by Strassler.\cite{strassler1,strassler2}
Also, multiloop diagrams with one-fermion-loop and multiple
propagator insertions has been cast into Bern-Kosower-type rule
by Schmidt and Schubert,
and they applied the rule to the calculation of
two-loop QED $\beta$ function.\cite{ss}
On the other hand, a quite different approach was developed by
Lam, where he showed that expressions similar to Bern-Kosower
rule can be obtained starting from the conventional Feynman parameter
formula in abelian gauge theories even beyond one-loop
order.\cite{lam} 

In this paper we refine the ideas in the above approaches from field
theory, and derive a full 
Bern-Kosower-type rule for scalar QED: 
we derive a set of rules for calculating
$S$-matrix elements for any processes at any order of the coupling
constant.
Also we clarify correspondence to the conventional Feynman rule.
(The method we show in this paper can straightforwardly be extended to
the case of spinor QED.)

The main idea is:
\begin{enumerate}
\item
Express a set of diagrams connected by gauge
transformation (see Fig.3 below) by a single worldline path-integral.
\item
Use gauge transformation (known as integration by parts
technique\cite{bk,strassler2}) 
to simplify calculation.
\end{enumerate}

For those unfamiliar with worldline path-integral formalism,
relation to the conventional
Feynman rule may be seen as follows.
Let us express the Feynman propagator in coordinate space 
using Feynman parameter\footnote{
Throughout the paper we work in $D$ dimensional space-time 
with the metric tensor
$g_{\mu \nu}=\mbox{diag}(+1,\underbrace{-1,\ldots ,-1}_{D-1})$.
}:
\begin{eqnarray}
i \Delta_F(x-y) &=& \int \frac{d^Dp}{(2\pi)^D} \, 
\frac{i \, e^{i p \cdot (x-y)}}{p^2-m^2+i\epsilon}
\label{fpfp1}
\\
&=& \int^\infty_0 d\alpha \, \int \frac{d^Dp}{(2\pi)^D} \, 
e^{ip \cdot (x-y) + i \alpha(p^2-m^2+i\epsilon)}
\label{fpfp2}
\\
&=& \int^\infty_0 d\alpha \, \,
i \left( \frac{1}{4\pi i\alpha} \right)^{D/2}
\exp \left[ - \, \frac{i}{4\alpha}(x-y)^2 - i\alpha(m^2-i\epsilon) \right] .
\label{fpfp3}
\end{eqnarray}
Note that (part of) the integrand in eqs.(\ref{fpfp2}) and
(\ref{fpfp3})
has a similar form to the propagator of a non-relativistic free
particle if $\alpha (>0)$ is identified with the time interval of propagation:
\begin{eqnarray}
K(x-y;\alpha) &\equiv & \int \frac{d^Dp}{(2\pi)^D} \, 
e^{ip \cdot (x-y) +i\alpha p^2}
\label{nrp1}
\\
&=& i \left( \frac{1}{4\pi i\alpha} \right)^{D/2}
\exp \left[ -\frac{i}{4\alpha}(x-y)^2 \right] .
\label{nrp2}
\end{eqnarray}
Namely, it satisfies 
\begin{eqnarray}
\left( i \frac{\partial}{\partial \alpha} - 
\frac{\partial}{\partial x^\mu}\frac{\partial}{\partial x_\mu} \right)
K(x-y;\alpha) = 0,
\\
K(x-y;+0) = \delta (x-y).
\label{propk2}
\end{eqnarray}
Hence, the associativity relation 
\begin{eqnarray}
\int d^D z \, \, K(x-z;\alpha_1) \, K(z-y;\alpha_2)
= K(x-y;\alpha_1+\alpha_2) 
\label{assoc}
\end{eqnarray}
holds as an important property of $K$ (see Fig.1), which
can be shown easily from eq.(\ref{nrp1}).
This property allows one to insert arbitrary number of
vertices along the
propagator lines of a given diagram, and 
if infinitely many are inserted, the integral expression
reduces to the path-integral.
%ppppppppppppppppppppppppppppppppppppppppppppppppppppppppppppp
%
\begin{figure}
\begin{center}
\epsfile{file=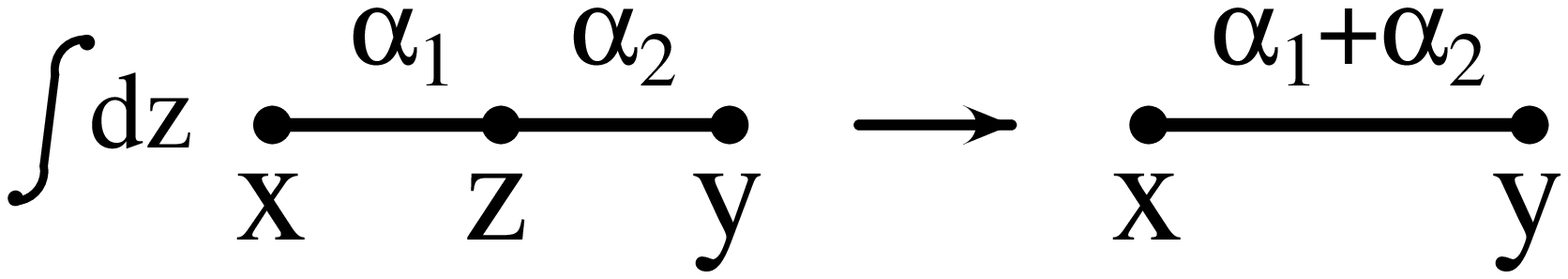,width=13cm}
\caption{A diagrammatical representation of the associativity
 relation satisfied by $K(x-y;\alpha)$.}
\end{center}
\end{figure}
%
%ppppppppppppppppppppppppppppppppppppppppppppppppppppppppppppp

In section 2, we derive the path-integral expression for a general set
of diagrams starting from quantum field theory, and derive
the general
expression after integration over $x(\tau)$.
Section 3 clarifies correspondence of the proper time integral formula
obtained in the previous section and the Feynman parameter integral
formula obtained from the conventional Feynman rule.
This enables one to express the two-point function 
(correlation function) 
$\left< x(\tau) x(\tau') \right>$
on the general diagram in terms of basic elements.
Section 4 explains a general prescription for integration by parts and
discuss relation to the gauge transformation on worldline.
The gauge-fixing parameter dependence of a set of diagrams is 
discussed in section
5.
The Bern-Kosower-type rule for a general set of diagrams is summarized
in section 6.
The rule for calculating a set of
diagrams including other than gauge interactions is demonstrated in
section 7.
Concluding remarks are given in section 8.

In Appendix A, details of calculation required in section 3 are shown.
Some properties of (counterpart of) the two-point function are 
listed in Appendix B with proofs.
A sample calculation using the Bern-Kosower-type
rule is shown in Appendix C.

\section{General Expression}

\cleqn

We consider scalar QED theory, whose Lagrangian is given by
\begin{eqnarray}
{\cal L}(\phi ,A_\mu ) = ( D_\mu \phi )^* ( D^\mu \phi )
- m^2 |\phi|^2 - \frac{\lambda}{4} |\phi|^4
- \frac{1}{4}F_{\mu\nu}F^{\mu\nu}
\end{eqnarray}
with
\begin{eqnarray}
D_\mu (A) = \partial_\mu - ie A_\mu (x).
\end{eqnarray}
We set $\lambda=0$ in most of the paper since the simplification of
calculation occurs regarding the gauge interactions.
The method for including $|\phi|^4$ interaction will be 
demonstrated in section 7.
As for the gauge-fixing term, we take Feynman gauge
\begin{eqnarray}
{\cal L}_{gf}(A_\mu) = - \frac{1}{2} ( \partial^\mu A_\mu )^2 
\end{eqnarray}
in the following, and discuss other gauge fixing conditions in section 5.

We start by defining a generating functional of 
connected Green functions,
which is {\it amputated} with respect to external
photons and {\it unamputated} with respect to external scalars:
\begin{eqnarray}
e^{W(J,J^* \! \! ,A_\mu)} &\equiv&
\left.
\int \! {\cal D}\phi \, {\cal D}Q_\mu \,
\exp \, {i \! \int dx \,
[ {\cal L}(\phi , Q_\mu)  + {\cal L}_{gf}(Q_\mu) 
+ J^*\phi + J\phi^* + j^\mu Q_\mu ]
} \,
\right|_{j_\mu \rightarrow - \Box A_\mu} ,
\end{eqnarray}
where $Q_\mu$ denotes quantum gauge field. 
Integrating out the scalar field, and then rewriting the integral
over $Q_\mu$ by functional derivatives, we obtain
\begin{eqnarray}
e^{W(J,J^* \! \! ,A_\mu)} &=&
\int {\cal D}Q_\mu \, \,
e^{ i \int dx \, 
[
\frac{1}{2} (A_\mu-Q_\mu) \Box (A^\mu-Q^\mu) 
- \frac{1}{2} A_\mu \Box A^\mu
]
}
\nonumber
\\ 
&& ~~~
\times
\exp [ \, {\textstyle 
- \mbox{Tr}\,\mbox{Log} ( D(Q)^2 + m^2 )
+ i \, \int \! \! \int dx dy \,
J^*(x) \left( 
\frac{1}{ D(Q)^2 + m^2 } 
\right)_{xy} \! J(y)
} ]
\\
&=& \rule{0mm}{9mm}
e^{- \frac{i}{2}\int dx A_\mu 
\Box
A^\mu}
\exp [ {\textstyle 
\frac{i}{2} \int \! \! \int dx dy
\frac{\delta}{\delta A_\mu(x)} (
{\hbox to 0pt{$\sqcap$}\sqcup}^{-1})_{xy}
\frac{\delta}{\delta A^\mu(y)} 
} ]
\nonumber
\\&& ~~~
\times
\exp [ \, {\textstyle
- \mbox{Tr}\,\mbox{Log} ( D(A)^2 + m^2 ) 
+ i \, \int \! \! \int dx dy \,
J^*(x) \left( \frac{1}{ D(A)^2 + m^2 } \right)_{xy}  \! J(y)
} ] ,
\label{deriv}
\end{eqnarray}
where we used functional analogue of an
identity\footnote{
To derive the integral form (left-hand-side) from the differential form
(right-hand-side), substitute
\begin{eqnarray}
f(\xi) = \int d\eta \, \delta (\xi - \eta) f(\eta)
= \int \frac{dp \, d\eta}{2\pi} \, e^{ip(\xi -\eta)}f(\eta)
\nonumber
\end{eqnarray}
and integrate over $p$ after replacing 
$d/d\xi$ by $ip$.
}
\begin{eqnarray}
\int^\infty_{-\infty} 
{\textstyle \frac{d\eta}{\sqrt{2\pi i a}} } \, \,
e^{i \frac{(\xi-\eta)^2}{2a} }
f(\eta)
= e^{ \frac{1}{2}ia \frac{d^2}{d\xi^2} } f(\xi) .
\end{eqnarray}

Interaction terms in eq.(\ref{deriv}), 
which functional derivatives operate, 
can be represented by path-integrals of a particle
interacting with the background
gauge field $A_\mu$, respectively, as
\begin{eqnarray}
- \mbox{Tr}\,\mbox{Log} ( D(A)^2 + m^2 ) 
&=& \int^\infty_0 \frac{dT}{T} \, e^{-im^2T}
{\hbox to 18pt{
\hbox to -3pt{$\displaystyle \int$} 
\raise-15pt\hbox{$\scriptstyle x(0)=x(T)$} 
}}
{\cal D}x(\tau) \,
\exp \biggl[
-i \int^T_0 d\tau \biggl(
\frac{\dot{x}^2}{4}-e A(x) \cdot \dot{x} 
\biggl) \biggl] ,
\label{pi1}
\\
{\textstyle 
-i \left( \frac{1}{ D(A)^2 + m^2 } \right)_{wz} 
}
&=&
\int^\infty_0 {dT} \, e^{-im^2T}
{\hbox to 18pt{
\hbox to -3pt{$\displaystyle \int$} 
\raise-15pt\hbox to 7pt{$\scriptstyle x(0)=z$} 
\raise18pt\hbox{$\scriptstyle x(T)=w \rule{0mm}{7mm}$}
}}
{\cal D}x(\tau) \,
\exp \biggl[
-i \int^{T}_0 d\tau \biggl(
\frac{\dot{x}^2}{4}-e A(x) \cdot \dot{x} 
\biggl) \biggl] .
\label{pi2}
\end{eqnarray}
Derivation of the first equation is given in Ref.\cite{strassler1},
and the second expression can be shown similarly.
The above interaction terms, respectively, 
correspond to a closed scalar chain (making
a loop) and an open scalar chain (whose both ends are connected to
external scalars) in the background gauge field.
Each term corresponds to the sum of Feynman diagrams with different
location of photons along the scalar chain, including arbitrary number
of  three-point
vertices and seagull vertices; see Fig.2.
Eq.(\ref{deriv}) has a simple form of connecting the
two kinds of scalar chains by photon propagators 
$i g_{\mu\nu}({\hbox to 0pt{$\sqcap$}\sqcup}^{-1})_{xy}$,
which serves for deriving path-integral expression for (a set of)
diagrams.
%ppppppppppppppppppppppppppppppppppppppppppppppppppppppppppppp
%
\begin{figure}
\begin{center}
\epsfile{file=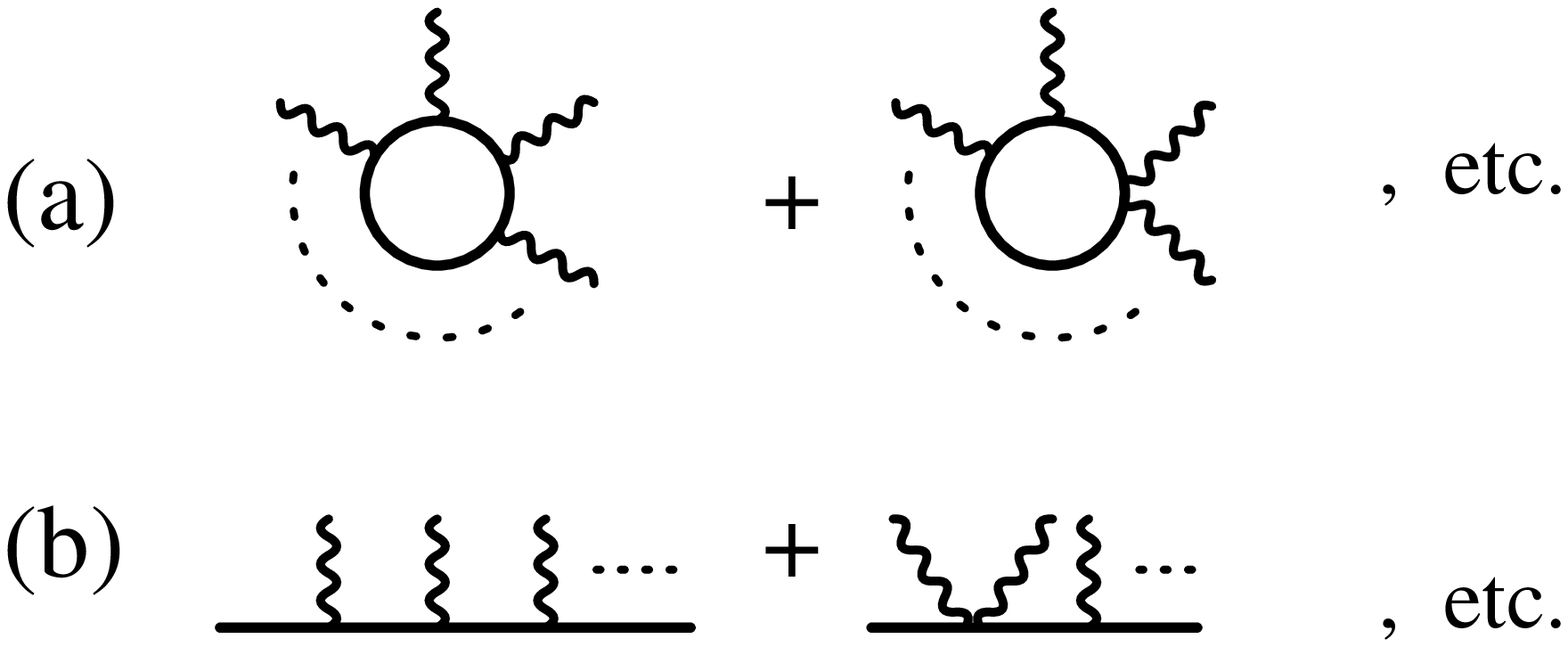,width=13cm}
\caption{The path-integral representation of a scalar particle 
interacting with the background gauge field 
a) where the scalar line is making a loop,
corresponding to eq.(\protect\ref{pi1}\protect), and 
b) where the scalar line is connected to external
lines,corresponding to eq.(\protect\ref{pi2}\protect).}
\end{center}
\end{figure}
%
%ppppppppppppppppppppppppppppppppppppppppppppppppppppppppppppp

Consider first a specific example.
We will find a convenient expression for the contribution of the set of
diagrams shown in Fig.3 (hereafter referred to as set I diagrams)
to the momentum space Green function defined by
\begin{eqnarray}
G(k_1,k_4;k_3,\epsilon_3,k_6,\epsilon_6) &\equiv&
\int dx dx' dw dz \, \,
e^{i (k_1 \cdot z + k_4 \cdot w + k_3 \cdot x + k_6 \cdot x') }
\nonumber \\ &&
\times
\left.
{\textstyle
\frac{\delta}{\delta J(z)} \,
\frac{\delta}{\delta J^*(w)} \,
\epsilon_{3\mu} \frac{\delta}{\delta A_\mu(x)} \,
\epsilon_{6\nu} \frac{\delta}{\delta A_\nu(x')}
} \,
W(J,J^* \! \! ,A)
\right|_{J=J^*=A=0} .
\label{defgreen}
\end{eqnarray}
All external momenta are taken to be outgoing.
%ppppppppppppppppppppppppppppppppppppppppppppppppppppppppppppp
%
\begin{figure}
\begin{center}
\epsfile{file=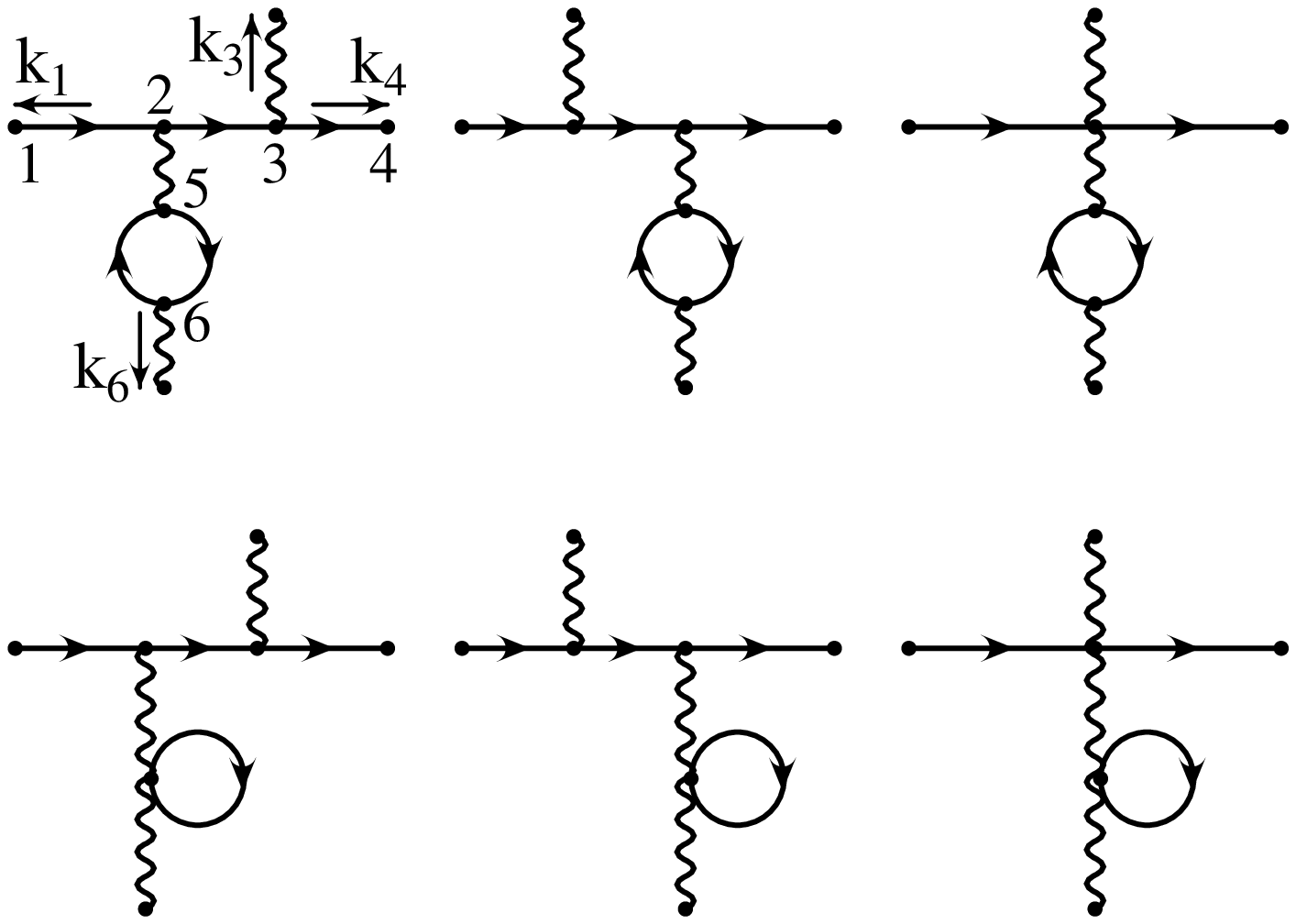,width=12cm}
\caption{The set I diagrams, which includes diagrams interrelated to 
one another
by the gauge transformation of internal and external photons.
}
\end{center}
\end{figure}
%
%ppppppppppppppppppppppppppppppppppppppppppppppppppppppppppppp

Let us choose the first diagram in the set I as the representative, 
and extract step by step the relevant terms in eq.(\ref{defgreen});
following procedure is 
sufficient for including all contributions from the set I 
diagrams.
After substituting (\ref{pi1}) and (\ref{pi2}) into (\ref{deriv}), 
we keep the term including one open scalar chain, one closed
scalar chain, and one internal photon propagator:
\begin{eqnarray}
W &\sim &
\frac{i}{2} \int \! \int dx dy \,
{\textstyle 
\frac{\delta}{\delta A_\mu(x)} (
{\hbox to 0pt{$\sqcap$}\sqcup}^{-1})_{xy}
\frac{\delta}{\delta A^\mu(y)} 
}
\nonumber
\\ &&
\times
{\hbox to 10pt{
\hbox to -3pt{$\displaystyle \int$} 
\raise-15pt\hbox to 7pt{$\scriptstyle 0 $} 
\raise18pt\hbox{$\scriptstyle \infty \rule{0mm}{4mm}$}
}}
dT
\, e^{-im^2T}
\int \! dw dz \, J^*(w) J(z)
{\hbox to 10pt{
\hbox to -3pt{$\displaystyle \int$} 
\raise-15pt\hbox to 7pt{$\scriptstyle z$} 
\raise18pt\hbox{$\scriptstyle w$}
}}
{\cal D}x \,
\exp \biggl[
-i 
{\hbox to 10pt{
\hbox to -3pt{$\displaystyle \int$} 
\raise-15pt\hbox to 7pt{$\scriptstyle 0 $} 
\raise18pt\hbox{$\scriptstyle T$}
}}
d\tau \, \left(
\frac{1}{4} \dot{x}^2 - e A(x) \cdot \dot{x} 
\right) 
\biggl] 
\nonumber
\\ &&
\times
{\hbox to 13pt{
\hbox to -3pt{$\displaystyle \int$} 
\raise-15pt\hbox to 4pt{$\scriptstyle 0$} 
\raise20pt\hbox{$\scriptstyle \infty \rule{0mm}{3mm}$}
}}
\frac{dT'}{T'} 
\, e^{-im^2T'}
\oint {\cal D}x' \,
\exp \biggl[
-i 
{\hbox to 10pt{
\hbox to -3pt{$\displaystyle \int$} 
\raise-15pt\hbox to 7pt{$\scriptstyle 0 $} 
\raise18pt\hbox{$\scriptstyle T'$}
}}
d\tau' \, \left(
\frac{1}{4} \dot{x}'^2 - e A(x') \cdot \dot{x}' 
\right) 
\biggl] .
\label{eg1}
\end{eqnarray}
We expand the integrand 
in powers of the coupling $e$, and extract the term corresponding to
two photon insertions in each scalar chain:
\begin{eqnarray}
\frac{(ie)^2}{2}
{\hbox to 10pt{
\hbox to -3pt{$\displaystyle \int$} 
\raise-15pt\hbox to 7pt{$\scriptstyle 0 $} 
\raise18pt\hbox{$\scriptstyle T$}
}}
dt_2 \, A(x_2) \cdot \dot{x}_2
{\hbox to 10pt{
\hbox to -3pt{$\displaystyle \int$} 
\raise-15pt\hbox to 7pt{$\scriptstyle 0 $} 
\raise18pt\hbox{$\scriptstyle T$}
}}
dt_3 \, A(x_3) \cdot \dot{x}_3
\times
\frac{(ie)^2}{2}
{\hbox to 10pt{
\hbox to -3pt{$\displaystyle \int$} 
\raise-15pt\hbox to 7pt{$\scriptstyle 0 $} 
\raise18pt\hbox{$\scriptstyle T'$}
}}
dt_5 \, A(x'_5) \cdot \dot{x}'_5
{\hbox to 10pt{
\hbox to -3pt{$\displaystyle \int$} 
\raise-15pt\hbox to 7pt{$\scriptstyle 0 $} 
\raise18pt\hbox{$\scriptstyle T'$}
}}
dt_2 \, A(x'_6) \cdot \dot{x}'_6
,
\label{connectpp}
\end{eqnarray}
where $x_i \equiv x(t_i)$ and $x'_j \equiv x'(t_j)$.
Then connect the internal photon propagator by taking derivative 
as
\begin{eqnarray}
\frac{\delta}{\delta A_\mu(x)} 
\frac{\delta}{\delta A^\mu(y)} ~
[ A(x_2) \cdot \dot{x}_2 ] \,
[ A(x'_5) \cdot \dot{x}'_5 ]
= \dot{x}_2 \cdot \dot{x}'_5 \, \,
[ \delta ( x_2 - x ) \delta ( x'_5 - y ) 
+ ( x \leftrightarrow y ) ] .
\end{eqnarray}
There are also terms in which $A(x_3)$ and $A(x_6')$ are
differentiated instead of 
$A(x_2)$ and $A(x_5')$, respectively,
so the factor $1/4$ in
(\ref{connectpp}) gets cancelled.
According to the definition (\ref{defgreen}),
the Green function is obtained by
substituting\footnote{
Note that 
in the case where $n$ external photon vertices are on some chain,
one should multiply by $n!$ after substituting 
$A^\mu (x(t_i)) = \epsilon_i^\mu e^{i k_i \cdot x(t_i)}$.
}
\begin{eqnarray}
&&
J^*(w) = e^{i k_4 \cdot w}, ~~~
J(z)= e^{i k_1 \cdot z}, ~~~
A^\mu (x_3) = \epsilon_3^\mu e^{i k_3 \cdot x_3} ,~~~
A^\mu (x'_6) = 
\epsilon_6^\mu e^{i k_6 \cdot x'_6}
\end{eqnarray}
to eq.(\ref{eg1}).
Thus,
\begin{eqnarray}
G_{\mbox{\scriptsize I}} (k,\epsilon) &=&
ie^4
\int dx dy \, 
({\hbox to 0pt{$\sqcap$}\sqcup}^{-1})_{xy}
{\hbox to 10pt{
\hbox to -3pt{$\displaystyle \int$} 
\raise-15pt\hbox to 7pt{$\scriptstyle 0 $} 
\raise18pt\hbox{$\scriptstyle \infty$}
}}
dT
\, e^{-im^2T}
{\hbox to 13pt{
\hbox to -3pt{$\displaystyle \int$} 
\raise-15pt\hbox to 4pt{$\scriptstyle 0$} 
\raise20pt\hbox{$\scriptstyle \infty \rule{0mm}{4mm}$}
}}
\frac{dT'}{T'} 
\, e^{-im^2T'}
\int^{T}_0 dt_2 dt_3 \int^{T'}_0 \! dt_5 dt_6
\nonumber
\\ &&
\times
\int dw dz 
{\hbox to 10pt{
\hbox to -3pt{$\displaystyle \int$} 
\raise-15pt\hbox to 7pt{$\scriptstyle z$} 
\raise18pt\hbox{$\scriptstyle w$}
}}
{\cal D}x \,
e^{
-i \int^{T}_0 \!
d\tau \, \frac{1}{4} \dot{x}^2
}
\oint {\cal D}x' \,
e^{
-i \int^{T'}_0 \!
d\tau' \, \frac{1}{4} \dot{x}'^2
}
\delta(x_2-x) \delta (x'_5-y)
\nonumber
\\ && ~~~~~
\times
e^{i (k_1 \cdot z + k_4 \cdot w ) }
\left( \dot{x}_2 \cdot \dot{x}'_5 \right)
\left( \epsilon_3 \cdot \dot{x}_3 e^{i k_3 \cdot x_3} \right)
\left( \epsilon_6 \cdot \dot{x}'_6 e^{i k_6 \cdot x'_6} \right)
\\
&=&
e^4
\int^\infty_0 d\alpha
{\hbox to 10pt{
\hbox to -3pt{$\displaystyle \int$} 
\raise-15pt\hbox to 7pt{$\scriptstyle 0 $} 
\raise18pt\hbox{$\scriptstyle \infty$}
}}
dT
\, e^{-im^2T}
{\hbox to 13pt{
\hbox to -3pt{$\displaystyle \int$} 
\raise-15pt\hbox to 4pt{$\scriptstyle 0$} 
\raise20pt\hbox{$\scriptstyle \infty \rule{0mm}{4mm}$}
}}
\frac{dT'}{T'} 
\, e^{-im^2T'}
\int^{T}_0 dt_2 dt_3 \int^{T'}_0 \! dt_5 dt_6
\nonumber
\\ &&
\times
\int_{\mbox{\scriptsize I}} {\cal D}x(\tau) \,
e^{
-i \int \!
d\tau \, \frac{1}{4} \dot{x}(\tau)^2
} \,
e^{i (k_1 \cdot z + k_4 \cdot w ) }
\left( - \dot{x}_2 \cdot \dot{x}'_5 \right)
\left( \epsilon_3 \cdot \dot{x}_3 e^{i k_3 \cdot x_3} \right)
\left( \epsilon_6 \cdot \dot{x}'_6 e^{i k_6 \cdot x'_6} \right)
,
\label{gi2}
\end{eqnarray}
where we have expressed the photon propagator using Feynman parameter,
and defined a ``path-integral over the 
set I diagrams''\footnote{
To be precise, we have expressed scalar chains in path-integrals and
photon propagators in Feynman parameter integrals.
} as
\begin{eqnarray}
\int_{\mbox{\scriptsize I}} {\cal D}x(\tau)
\, e^{-i\int d\tau \frac{1}{4} \dot{x}(\tau)^2}
&\equiv&
\int dx dy \, 
i \left( \frac{1}{4\pi i\alpha} \right)^{D/2}
e^{-\frac{i}{4\alpha} (x-y)^2}
\nonumber 
\\&& \times
\int dw dz
{\hbox to 10pt{
\hbox to -3pt{$\displaystyle \int$} 
\raise-15pt\hbox to 7pt{$\scriptstyle z$} 
\raise18pt\hbox{$\scriptstyle w$}
}}
{\cal D}x \,
e^{ -i 
\int^{T}_0 \! d\tau \, 
\frac{1}{4} \dot{x}^2 
} 
\oint {\cal D}x' \,
e^{
-i \int^{T'}_0 \! d\tau' \, \frac{1}{4} \dot{x}'^2 
}
\nonumber
\\&& ~~~~~~~
\times 
\delta(x_2-x) \delta (x'_5-y) .
\label{piod}
\end{eqnarray}

Since the path-integral over $x(\tau)$ is Gaussian, it is
straightforward (at least formally) to perform the integration.
For convenience, we assign an outgoing momentum $k_i$ 
and a polarization vector $\epsilon_i$
to every vertex ($x_1 \equiv z$, $x_4 \equiv w$),
and replace the vertex factors by an exponential factor:
\begin{eqnarray}
e^{i (k_1 \cdot z + k_4 \cdot w ) }
\left( - \dot{x}_2 \cdot \dot{x}'_5 \right)
\left( \epsilon_3 \cdot \dot{x}_3 e^{i k_3 \cdot x_3} \right)
\left( \epsilon_6 \cdot \dot{x}'_6 e^{i k_6 \cdot x'_6} \right)
\longrightarrow
\exp \left[ 
\sum_{i=1}^{6} (i k_i \cdot x_i + \epsilon_i \cdot \dot{x}_i)
\right] .
\label{exponentiate}
\end{eqnarray}
At the end of the calculation, to
recover the correct result:
\begin{enumerate}
\renewcommand{\labelenumi}{\arabic{enumi})}
\item 
We set
$k_2 = k_5 =0$ and $\epsilon_1 = \epsilon_4 = 0$. 
\item
Only the terms in which each polarizatoin vector 
$\epsilon_2,\epsilon_3,\epsilon_5,\epsilon_6$
appears precisely once
(multi-linear in each polarization vector)
are retained.
\item
We replace the internal photon wave function as
\begin{eqnarray}
\epsilon_2^\mu \epsilon_5^\nu \rightarrow -g^{\mu \nu} .
\end{eqnarray} 
\end{enumerate}
The replacement (\ref{exponentiate}) simplifies the
integration over $x(\tau)$.
Hence, we obtain
\begin{eqnarray}
G_{\mbox{\scriptsize I}} (k,\epsilon) &=&
 e^4
\int^\infty_0 d\alpha
{\hbox to 10pt{
\hbox to -3pt{$\displaystyle \int$} 
\raise-15pt\hbox to 7pt{$\scriptstyle 0 $} 
\raise18pt\hbox{$\scriptstyle \infty$}
}}
dT
\, e^{-im^2T}
{\hbox to 13pt{
\hbox to -3pt{$\displaystyle \int$} 
\raise-15pt\hbox to 4pt{$\scriptstyle 0$} 
\raise20pt\hbox{$\scriptstyle \infty \rule{0mm}{4mm}$}
}}
\frac{dT'}{T'} 
\, e^{-im^2T'}
\int^{T}_0 dt_2 dt_3 \int^{T'}_0 \! dt_5 dt_6
\nonumber
\\ &&
\times
{\cal N} \, \exp \biggl[ \, \frac{1}{2}
\sum_{i,j=1}^6 \biggl\{
- i k_i \cdot k_j G_B^{ij}
- 2 k_i \cdot \epsilon_j \partial_j G_B^{ij}
+ i \epsilon_i \cdot \epsilon_j \partial_i \partial_j G_B^{ij}
\biggl\}
\biggl] , 
\label{egfinal}
\end{eqnarray}
where the normalization factor is defined by
\begin{eqnarray}
{\cal N} \equiv \int_{\mbox{\scriptsize I}} {\cal D}x(\tau)
\, e^{-i\int d\tau \frac{1}{4} \dot{x}(\tau)^2} ,
\label{egnorm}
\end{eqnarray}
and the two-point functions are given by
\begin{eqnarray}
\begin{array}{lcc}
g^{\mu \nu} \, G_B^{ij} &=& -i \left< x^\mu (t_i) x^\nu (t_j) \right>,
\\
\rule{0mm}{7mm}
g^{\mu \nu} \, \partial_j G_B^{ij} &=& -i 
\left< x^\mu (t_i) \dot{x}^\nu (t_j) \right>,
\\
\rule{0mm}{7mm}
g^{\mu \nu} \, \partial_i \partial_j G_B^{ij} 
&=& -i \left< \dot{x}^\mu (t_i) \dot{x}^\nu (t_j) \right>,
\end{array}
\end{eqnarray}
with the expectation value
taken with respect to the path-integral average over the set I diagrams:
\begin{eqnarray}
\left< {\cal O}(x) \right> \equiv {\cal N}^{-1}
\int_{\mbox{\scriptsize I}} {\cal D}x(\tau) \, {\cal O} (x)
\, e^{-i\int d\tau \frac{1}{4} \dot{x}(\tau)^2} .
\label{egexpect}
\end{eqnarray}
We remind the reader that $\partial_j G_B^{ij}$ {\it differs} from the
differentiation of $G_B^{ij}$ with respect to $t_j$.
Precise definition will be made clear in the next section.

So far we considered a specific example.
The steps that led to eq.(\ref{egfinal}) can be generalized to an
arbitrary set of diagrams:
A set of diagrams consists of those which can be transformed to one
another by sliding photon legs along the scalar chains, where any two
three-point vertices on a same chain may combine to become a seagull
vertex.
Any single set contains all diagrams that are interrelated to 
one another by the gauge transformation of external and internal
photons. 
In other words, each set constitutes a gauge-invariant subamplitude if
the external scalar propagators are amputated and taken to be on-shell,
$k_s^2 \rightarrow m^2$.\footnote{
This is true only for the renormalized Green function.
}
Thus, the Green function
\begin{eqnarray}
G(k,\epsilon) &=&
\int \prod_i dx_i \, e^{i \sum k_i \cdot x_i}
\left[
\prod {\textstyle \frac{\delta}{i\delta J(w_i)} } \,
\prod
{\textstyle \frac{\delta}{i\delta J^*(z_i)} } \,
\prod {\textstyle \epsilon_i^\mu 
\frac{\delta}{i\delta A^\mu(y_i)} } \, \,
W(J,J^* \! \! ,A)
\right]_{J=J^*=A=0} 
\label{defg}
\end{eqnarray}
at each order of the coupling $e$ can be
decomposed to the sub-Green functions corresponding to the sets $S$ of
diagrams as
\begin{eqnarray}
G(k,\epsilon ) = \sum_S G_S (k,\epsilon) ,
\end{eqnarray}
where the decomposition is accomplished naturally
by expanding eq.(\ref{deriv}) in 
powers of $e$, taking functional derivatives, and then substituting the
external wave functions; see eqs.(\ref{eg1})-(\ref{gi2}).

Following similar steps as in the former example, it is easy to see that 
the sub-Green function for a set $S$ with $2n_s$
external scalars at ${\cal O}(e^n)$ is given generally by
\begin{eqnarray}
G_S(k,\epsilon) &=& (ie)^n \, C \,
\int^\infty_0 \prod_r d\alpha_r \, \prod_{chain \, l}
\left(
\int^\infty_0 [dT_l]\, e^{-im^2T_l} \int^{T_l}_0 \prod_{i_l} dt_{i_l}
\right)
\nonumber
\\ &&
\times
{\cal N} \, \exp \biggl[ \, \frac{1}{2}
\sum_{i,j=1}^{n+2n_s} \biggl\{
- i k_i \cdot k_j G_B^{ij}
- 2 k_i \cdot \epsilon_j \partial_j G_B^{ij}
+ i \epsilon_i \cdot \epsilon_j \partial_i \partial_j G_B^{ij}
\biggl\}
\biggl] , 
\label{generalexp}
\end{eqnarray}
where $C$ is the combinatorial factor\footnote{
The combinatorial factor $C$ in general differs from 
$(\mbox{symmetry factor}) \times (\mbox{statistical factor})$ of the
corresponding Feynman diagrams, since certain diagrams do not
distinguish the interchange of photon legs.
e.g.\ $C=1/2$ for the scalar self-energy at one-loop.
}, 
$\alpha_r$ denotes the Feynman
parameter of the $r$-th photon propagator.
The chain $l$ represents
open or closed scalar chain, 
and the integral measure for its length
$T_l$ is 
\begin{eqnarray}
[dT_l] = \left\{
\begin{array}{cl}
dT_l & \mbox{for $l=$open}\\
\rule{0mm}{4mm} {dT_l}/{T_l} & \mbox{for $l=$closed}
\end{array} 
\right. .
\label{imeasure}
\end{eqnarray}
$i_l$ represents photon vertex on the chain $l$.
For convenience, we assigned an outgoing external momentum $k_i$ and a
polarization vector $\epsilon_i$ to every vertex $i$. 
Normalization factor $\cal N$ and two-point functions 
$G_B^{ij}$, $\partial_j G_B^{ij}$, and
$\partial_i \partial_j G_B^{ij}$
are defined similarly as eqs.(\ref{egnorm})-(\ref{egexpect}), but for
the path-integral over the set $S$ diagrams.
The exponential factor is
common to all
$S$ once the numbers of external scalars and 
photons as well as the order of
$e$ are fixed.
(Explicit forms of $G_B^{ij}$'s depend on $S$, though.)

Furthermore, one should manipulate following processes (dependent on the
set $S$) to the above
$G_S(k,\epsilon)$: 
\begin{enumerate}
\renewcommand{\labelenumi}{\arabic{enumi})}
\item 
If the vertex $i$ is internal, we set corresponding $k_i=0$. 
\item
If the vertex $i$ is an endpoint of an open scalar chain,
we set corresponding $\epsilon_i=0$.
\item
Only the terms multi-linear in each remaining polarization vector
are kept.
\item
We replace the polarization vectors at both ends ($i_r$ and $j_r$) 
of every photon
propagator $r$ as
\begin{eqnarray}
\epsilon_{i_r}^\mu \epsilon_{j_r}^\nu \rightarrow -g^{\mu \nu} .
\label{replpolv}
\end{eqnarray} 
\end{enumerate}

At this stage, one could directly evaluate the integrals in
eq.(\ref{generalexp}) once the explicit forms of $\cal N$ and 
$G_B^{ij}$, $\partial_j G_B^{ij}$, and
$\partial_i \partial_j G_B^{ij}$
are known.
It already has advantages that a set of diagrams is cast into
one single expression, and that the expressions for different sets of
diagrams can be obtained in 
similar simple manners.
Also, the spinor helicity technique \cite{sht1,sht2} can be used, so 
the number of independent 
dot products in the exponent can be reduced.
Moreover, the Bern-Kosower-type rule allows
use of partial integration technique, which simplify the calculation 
further.
After that, one will 
integrate over $\alpha_r$, $t_i$, and $T_l$.

In order to understand the remaining part of the rule, one needs a
close study of the two-point function 
\begin{eqnarray}
g^{\mu\nu} \, G_B(\tau , \tau') \equiv
-i
\left< x^\mu(\tau) x^\nu(\tau') \right>.
\label{deftf}
\end{eqnarray}
In principle, $G_B(\tau , \tau')$ is obtained by solving
\begin{eqnarray}
\frac{\partial^2}{\partial \tau^2} \, G_B(\tau , \tau') 
= 2 \, \delta (\tau -\tau')
\end{eqnarray}
after removing the zero mode,
where appropriate boundary condition should be imposed
at each internal vertex of the
diagram\cite{ss}.
We take, however, an alternative approach.
It is possible to find simple rules to express $G_B(\tau , \tau')$ 
for a general diagram in terms of basic elements.

\section{Relation to Feynman Parameter Formula and 
Decomposition of $G_B$}

\cleqn

In this section, we derive the Feynman parameter formula for a scalar
QED diagram (rather than for a set of diagrams considered in the
previous section).
In this formula a matrix $Z_{ij}$ appears, which is identified to be
the counterpart of $G_B^{ij}$.
$Z_{ij}$ is defined through integral 
over finite number of variables instead of
the path-integral formulation, which enables us to investigate its
properties in an unambiguous way.
We deal with a general $\phi^3$ diagram in subsection 3.a, followed by an
extension to scalar QED diagrams in subsection 3.b.
Then subsection 3.c will clarify the relation 
between the Feynman parameter integral 
formula and the general expression for $G_S (k,\epsilon)$
obtained in the last section.
Finally, we show how to decompose $\cal N$ and $G_B^{ij}$ 
to simpler elements
in subsection 3.d.

\subsection{Scalar $\phi^3$ Diagram}

For the calculation of a general $\phi^3$ diagram, it has long been 
known how to write down the Feynman parameter formula\cite{ll}. 
We rederive the formula in a manner convenient for application to 
the case of scalar QED diagram.

A general connected $\phi^3$ diagram 
with $n$ vertices and $N$ internal lines
can be written using
Feynman rule in coordinate space as
\begin{eqnarray}
i T = (ie)^n \int \prod^n_{i=1} d^D x_i \, e^{i \sum_i k_i \cdot x_i}
\left[ \prod^N_{r=1} i \Delta_F(x_{i_r}-x_{j_r}) \right],
\end{eqnarray}
where $e$ is the $\phi^3$ coupling constant.
$i_r$ and $j_r$ represent the vertices at both ends of the $r$-th internal
line.
For convenience an outgoing
external momentum $k_i$ is assigned to every vertex.
If the vertex is internal, we set the corresponding $k_i=0$ at the end
of the calculation.
Combinatorial factor, if any, is suppressed for simplicity.

Substituting the propagator given in eq.(\ref{fpfp3}), we have
\begin{eqnarray}
iT = (ie)^n \int^\infty_0 \prod^N_{r=1} d\alpha_r \,
e^{-i(m^2-i\epsilon )\sum_r \alpha_r} \, I(\alpha ) ,
\end{eqnarray}
where
%\footnote{
%One recognizes that if infinitely many vertices are inserted to
%(\ref{ialpha1}) 
%using the associativity relation (\ref{assoc}), the path-integral over
%the diagram, e.g.\ (\ref{piod}), is obtained.
%}
\begin{eqnarray}
I(\alpha ) \equiv \int [dx_i]
\, 
\exp \left[ - \frac{i}{4} \sum^n_{i,j=1} x_i \cdot x_j \, 
A_{ij}(\alpha ) + i \sum^n_{i=1} k_i \cdot x_i \right] ,
\label{ialpha1}
\end{eqnarray}
and
\begin{eqnarray}
\sum^n_{i,j=1} x_i \cdot x_j \, 
A_{ij}(\alpha ) \equiv \sum^N_{r=1} \frac{(x_{i_r}-x_{j_r})^2}{\alpha_r} .
\label{quad}
\end{eqnarray}
The matrix $A_{ij}(\alpha )$ represents the
topoplogy of the diagram (how the vertices are
connected).
We have absorbed the factor before exponential in eq.(\ref{nrp2}) into
the integral measure:
\begin{eqnarray}
[dx_i] \equiv
\left[ \prod^N_{r=1} i \left( \frac{1}{4\pi i\alpha_r}
\right)^{D/2} \right] \cdot
\prod^n_{i=1} d^D x_i .
\label{measure}
\end{eqnarray}
Note that it depends on Feynman parameters.

Then, after Gaussian integration over $x_i$'s in $I(\alpha)$, we will be 
left with the desired Feynman parameter integral formula.
Reflecting the invariance of the quadratic form (\ref{quad}) under 
translation
\begin{eqnarray}
x^\mu_i \rightarrow x^\mu_i + c^\mu ,
\label{transl}
\end{eqnarray}
the matrix $A_{ij}(\alpha )$ has a zero eigenvalue.
Namely, $I(\alpha)$ will be proportional to the $\delta$-function 
representing 
momentum conservation.
Indeed, after integration over $x_i$'s, we obtain
\begin{eqnarray}
I(\alpha)= 
(2\pi)^D \delta \left( \sum^n_{i=1} k_i \right) \cdot
i^l \left( \frac{1}{4\pi i}
\right)^{Dl/2} 
\Delta(\alpha)^{-D/2} \,
\exp \left[ i \sum^n_{i,j=1} k_i \cdot k_j Z_{ij}(\alpha) \right]
\label{ialpha}
\end{eqnarray}
with
\begin{eqnarray}
\Delta(\alpha) = \frac{1}{n} 
\left( \prod^N_{r=1} \alpha_r \right) 
\mbox{det}' A(\alpha) .
\label{delta}
\end{eqnarray}
Here, $l=N-n+1$ is the number of loop of the diagram.
$\det '$ denotes the product of eigenvalues but zero.
$Z_{ij}(\alpha)$ is the inverse of $A_{ij}(\alpha)$ after the zero mode
is removed, or, fixing the center of gravity of vertices.
Derivation of eqs.(\ref{ialpha}) and (\ref{delta}) is given in
Appendix A.

In eq.(\ref{ialpha}), $Z_{ij}(\alpha)$ is not uniquely determined.
This is because one can readily
confirm the invariance of $I(\alpha)$ under the
transformation of $Z$,
\begin{eqnarray}
Z_{ij}(\alpha) \rightarrow
Z_{ij}(\alpha) + f_i(\alpha) + f_j(\alpha)
~~~~~
\mbox{for $\forall f_i(\alpha)$}, 
\label{transfz}
\end{eqnarray}
due to the momentum
conservation.
Among the class of $Z(\alpha)$'s connected by the transformation, there
is a specific choice of $Z(\alpha)$ most convenient to the
following argument.
We choose
\begin{eqnarray}
g^{\mu\nu} Z_{ij}(\alpha) \equiv
- \frac{i}{4} 
\left< \! \left< \, (x_i-x_j)^\mu (x_i-x_j)^\nu \, \right> \! \right>
\label{defz}
\end{eqnarray}
with $\left< \! \left< \ldots \right> \! \right>$ defined by
\begin{eqnarray}
\left< \! \left< {\cal O} \right> \! \right> \equiv
\frac{ 
\int [dx_i] \, {\cal O} \, \exp [{-\frac{i}{4}\sum x_i \cdot x_j
A_{ij}}]
}
{
\int [dx_i] \, \exp [{-\frac{i}{4}\sum x_i \cdot x_j
A_{ij}}]
} .
\label{expect}
\end{eqnarray}
The numerator and the
denominator of eq.(\ref{expect}), respectively, are
ill-defined due to the zero eigenvalue of $A(\alpha)$, so one has to
first remove the zero mode in the integrals.
Because $x_i-x_j$ in eq.(\ref{defz}) is invariant under the translation
(\ref{transl}),
$Z(\alpha)$ thus defined is independent of how one removes the zero
mode.\footnote{
Naively, $Z(\alpha)$ being the inverse of $A(\alpha)$, one may consider
a natural definition would be 
$g^{\mu\nu} Z'_{ij}(\alpha) \equiv 
\frac{i}{2} \left< \! \left< x_i^\mu x_j^\nu \right> \! \right>$.
$Z'$ and $Z$ given by eq.(\ref{defz}) are equivalent under 
the transformation (\ref{transfz}) with $f_i = -Z'_{ii}/2$.
The disadvantage of $Z'$ is that it depends on how one removes the
zero mode in calculating 
$\left< \! \left< x_i^\mu x_j^\nu \right> \! \right>$ since $x_i^\mu x_j^\nu$
is not translationally invariant.
}
Lam has pointed out\cite{lam} that this choice of
$Z(\alpha)$ is characterized by the condition
\begin{eqnarray}
Z_{ii}(\alpha) = 0
~~~~~~~
\mbox{for $1 \leq i \leq n$},
\end{eqnarray}
and is called zero-diagonal level scheme.

We list some important properties of $Z_{ij}$ together with their proofs 
in Appendix B.

\subsection{Scalar QED Diagram}

Now we derive the Feynman parameter intergral formula
for a scalar QED diagram.
We consider diagrams contributing to the
Green function (\ref{defg}) which is amputated with respect to external
photons and unamputated with respect to external scalars.
%ppppppppppppppppppppppppppppppppppppppppppppppppppppppppppppp
%
\begin{figure}
\begin{center}
\epsfile{file=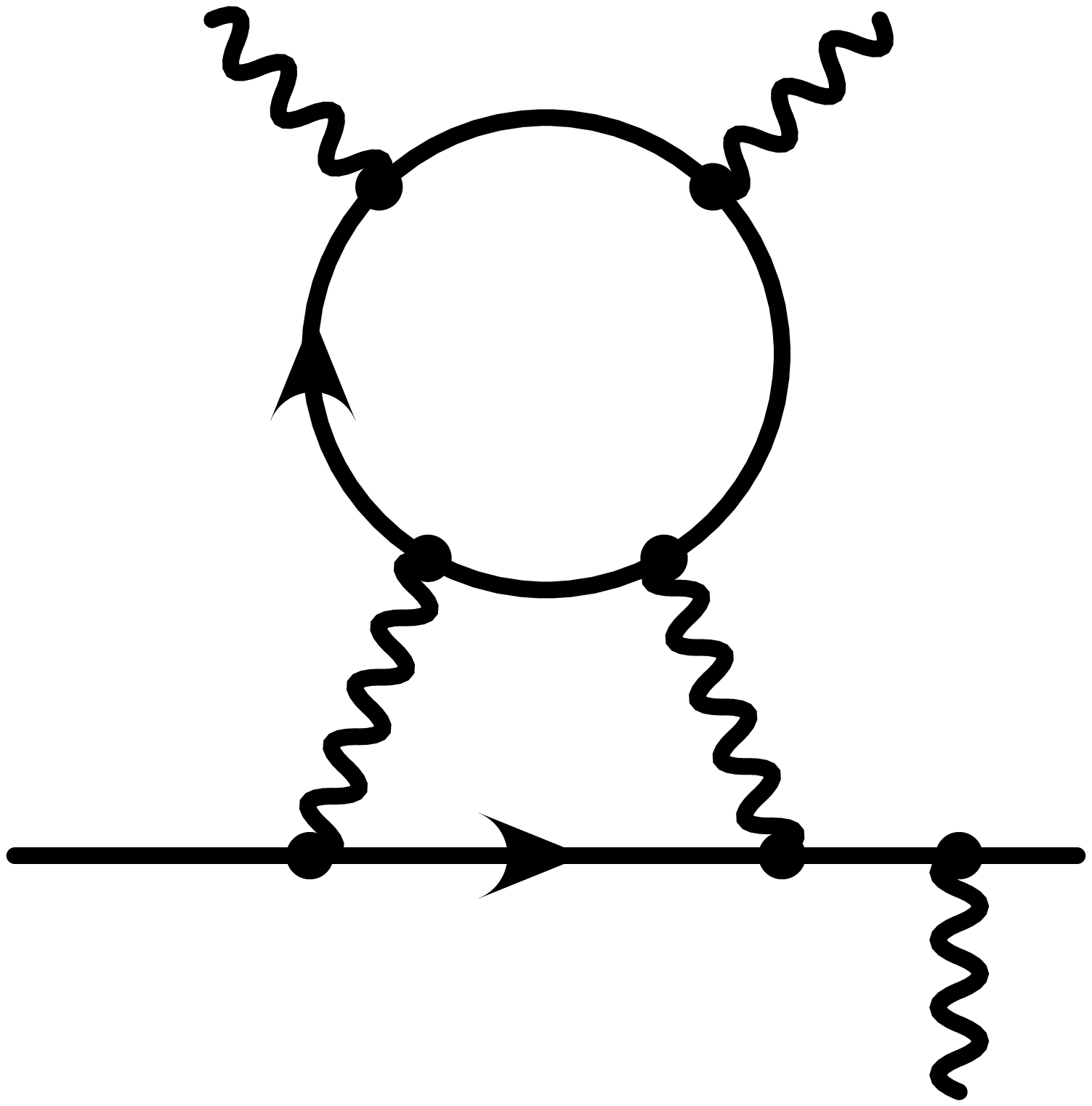,width=5cm}
\caption{A scalar QED diagram including only three-point gauge 
vertices, which
contributes to the Green function amputated with respect to external
photons and unamputated with respect to external scalars.}
\end{center}
\end{figure}
%
%ppppppppppppppppppppppppppppppppppppppppppppppppppppppppppppp

First, let us consider a diagram without seagull vertex; see Fig.4:
\begin{eqnarray}
G_D(k,\epsilon) 
= (ie)^n \int \prod_i d^Dx_i \, e^{i\sum k_i \cdot x_i}
\left[ \prod_{chain \, l} 
\left\{ \prod_{i_l=1}^{n_l}
i\Delta_F(x_{i_l+1}-x_{i_l})
\stackrel{\textstyle \leftrightarrow}{V_{i_l}}
\right\} \right] \,
&&
\nonumber
\\
\times
\prod_{photon \, r} i\Delta_F(x_{i_r}-x_{j_r})
&&
\left.
\rule{0mm}{10mm}
\right|
_{\textstyle \epsilon_{i_r}^\mu \epsilon_{j_r}^\nu \rightarrow -
g^{\mu\nu}} ,
\label{fundeq}
\end{eqnarray}
with the vertex operator
\begin{eqnarray}
\stackrel{\textstyle \leftrightarrow}{V_{j}} \equiv
\epsilon_{j}^\mu 
\left( i 
\stackrel{\textstyle \rightarrow}{\frac{\partial}{\partial x_{j}^\mu}}
-
i\stackrel{\textstyle \leftarrow}{\frac{\partial}{\partial x_{j}^\mu}}
\right) .
\end{eqnarray}
Here,
$i_l$'s ($1 \leq i_l \leq n_l$)
denote vertices on the 
scalar propagator
chain $l$, labelled in increasing order
along the charge flow on that chain.
For an open chain we suppressed one additional scalar propagator 
$i \Delta_F(x_1-x_0)$ on the right of the vertex operator 
$\stackrel{\textstyle \leftrightarrow}{V_1}$
in eq.(\ref{fundeq}).
$i_r$ and $j_r$
represent the vertices at both ends of the photon propagator $r$.
Again, we assign an outgoing external momentum $k_i$ and a
polarization vector $\epsilon_i$ to every vertex $i$. 
At the end of the calculation, we set $k_i=0$ for internal vertices, 
$\epsilon_i=0$ at the endpoints of open scalar chains, and 
also replace the polarization vectors at both ends of every internal 
photon line as 
$\epsilon_{i_r}^\mu \epsilon_{j_r}^\nu \rightarrow - g^{\mu\nu}$
(corresponding to taking Feynman gauge for photon propagator);
see Fig.5.
%ppppppppppppppppppppppppppppppppppppppppppppppppppppppppppppp
%
\begin{figure}
\begin{center}
\epsfile{file=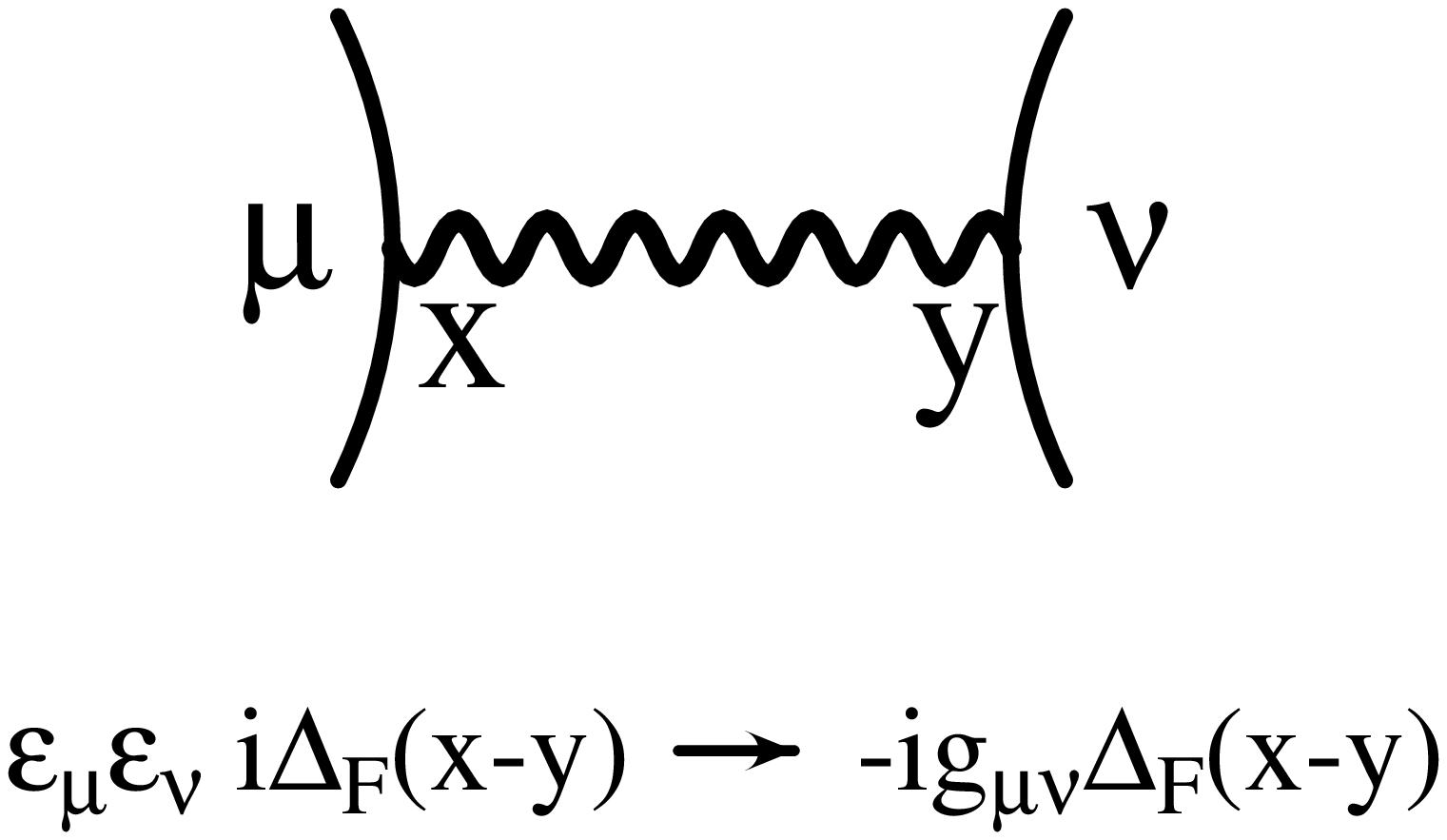,width=10cm}
\caption{The Feynman gauge photon propagator is obtained by replacing
internal photon polarization vectors at both ends of every photon
propagator 
by $-g_{\mu \nu}$.}
\end{center}
\end{figure}
%
%ppppppppppppppppppppppppppppppppppppppppppppppppppppppppppppp

Introducing Feynman parameter for every
propagator,
we have
\begin{eqnarray}
G_D(k,\epsilon) = (ie)^n 
\prod_l
\left(
\int^\infty_0 \prod_{i_l} d\alpha_{i_l}
\right)
\int^\infty_0 \prod_r d\alpha_r 
\, e^{-i\sum_l{T_l(m^2-i\epsilon)}} I(\alpha ),
\label{gd}
\end{eqnarray}
where
\begin{eqnarray}
I(\alpha ) 
\equiv 
\int \prod_i d^Dx_i \, 
e^{i\sum k_i \cdot x_i} 
\left[ \prod_l \left\{ \prod_{i_l}
K(x_{i_l+1}-x_{i_l};\alpha_{i_l})
\stackrel{\textstyle \leftrightarrow}{V_{i_l}}
\right\} \right] \,
\prod_r K(x_{i_r}-x_{j_r};\alpha_r).
\end{eqnarray}
$K$ is the propagator defined in eq.(\ref{nrp2});
$\alpha_{i_l}$ is the Feynman parameter between the vertices $i_l$ and
$i_l-1$, 
and $T_l = \sum_{i_l} \alpha_{i_l}$.
%ppppppppppppppppppppppppppppppppppppppppppppppppppppppppppppp
%
\begin{figure}
\begin{center}
\epsfile{file=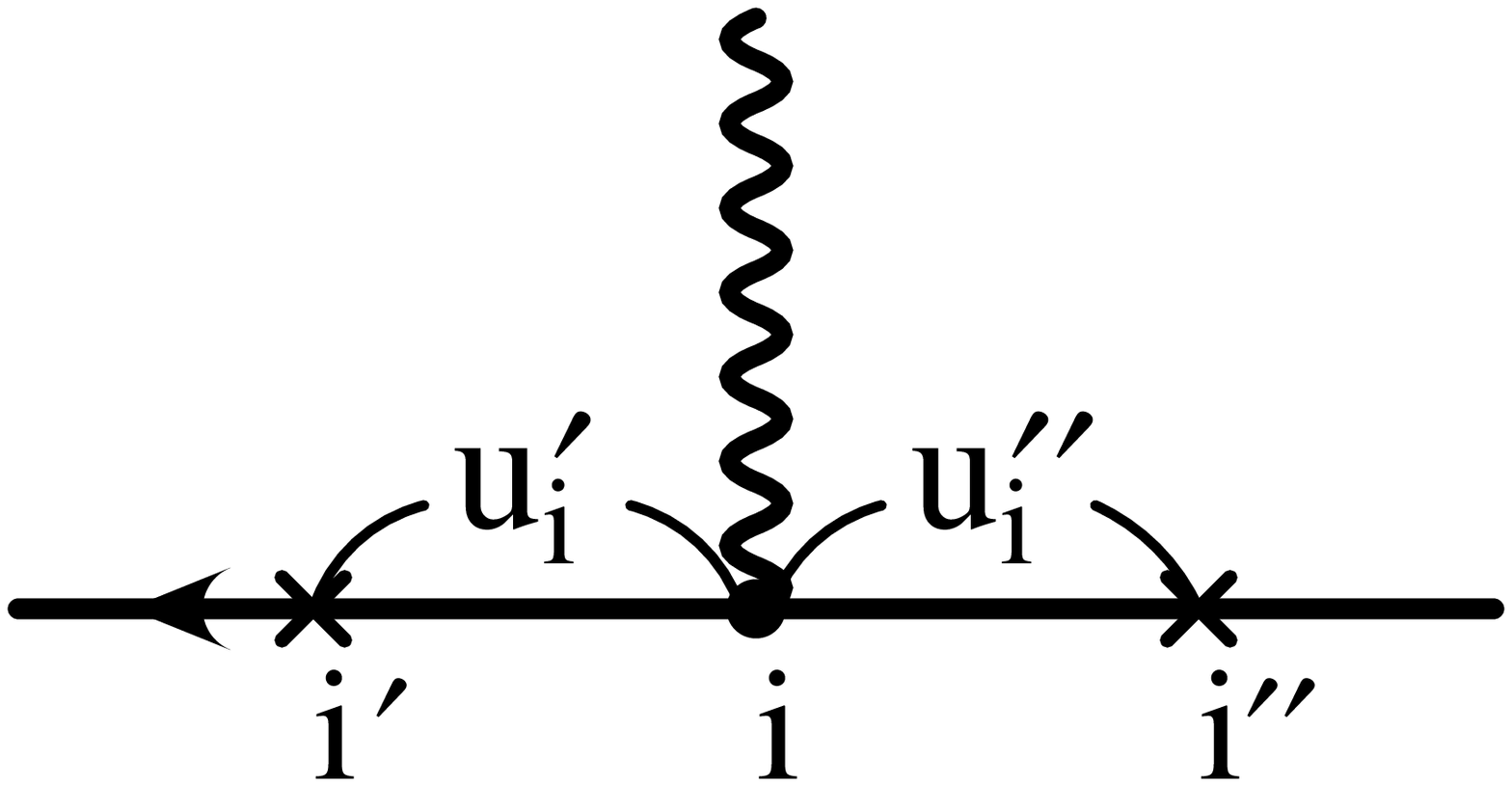,width=10cm}
\caption{The dummy vertices $i'$ and $i''$ inserted on 
both sides of every vertex
$i$ in the order $i''<i<i'$ along the charge flow on the scalar line.
The Feynman parameter between vertices $i'$ and $i$ ($i$ and $i''$) is
denoted as $u_i'$ ($u_i''$).}
\end{center}
\end{figure}
%
%ppppppppppppppppppppppppppppppppppppppppppppppppppppppppppppp

Before integrating over $x_i$'s in $I(\alpha )$, 
we would like to replace the vertex
operator 
$\stackrel{\textstyle \leftrightarrow}{V_{i}}$
by some simple factor {\it associated with the 
vertex} $i$.
To this end, we insert, on both sides
of every vertex $i$, dummy vertices 
$i'$ and $i''$ on the scalar line in the order 
$i'' < i < i'$
using the associativity relation (\ref{assoc}); see Fig.6.
Then we can replace the vertex operators acting on scalar propagators
as
\begin{eqnarray}
\stackrel{\textstyle \leftrightarrow}{V_{i}} 
~\longrightarrow ~
\frac{1}{2} \epsilon_i \cdot
\left( {\textstyle
\frac{x_i'-x_i}{u_i'} +
\frac{x_i-x_i''}{u_i''} 
}
\right) .
\end{eqnarray}
Hence, we have
\begin{eqnarray}
I(\alpha ) &=&
\int [dx_a] \,
\prod_i
\frac{1}{2} \epsilon_i \cdot
\left( {\textstyle
\frac{x_i'-x_i}{u_i'} +
\frac{x_i-x_i''}{u_i''} 
}
\right) \,
\exp \biggl[
-\frac{i}{4} \sum_{a,b} x_a \cdot x_b A_{ab}(\alpha ,u',u'') +
i \sum_i k_i \cdot x_i
\biggl] .
\label{qedialpha0}
\end{eqnarray}
Here, $a,b$ denote vertices including dummy vertices ($i,i'$, and
$i''$).
The matrix $A_{ab}$ and the measure $[dx_a]$, 
respectively, are defined similarly as in 
eqs.(\ref{quad}) and (\ref{measure}), but depend also on $u'$ and
$u''$. 
Note that $I(\alpha )$ is independent of $u_i'$ and $u''_i$, since
it is completely arbitrary where to insert dummy vertices 
as long as the order $i'' < i < i'$ is preserved.

To perform Gaussian integration over $x_a$'s, we exponentiate the
polarization vectors as in eq.(\ref{exponentiate}).
Defining a source
\begin{eqnarray}
J_a^\mu = \sum_i \left[
k_i^\mu \delta_{ia} - \frac{i}{2}\epsilon_i^\mu
\left( {\textstyle
\frac{\delta_{i'a}-\delta_{ia}}{u_i'} +
\frac{\delta_{ia}-\delta_{i''a}}{u_i''} 
}
\right)
\right] ,
\end{eqnarray}
we have
\begin{eqnarray}
I(\alpha ) &=&
\int [dx_a] \,
\exp \biggl[
-\frac{i}{4} \sum_{a,b} x_a \cdot x_b A_{ab}(\alpha ,u',u'') +
i \sum_a J_a \cdot x_a
\biggl]_{\mbox{linear in each $\epsilon$}}
\label{qedialpha1}
\\
&=&
(2\pi)^D \delta \left( \sum^n_{i=1} k_i \right) \cdot
i^l \left( \frac{1}{4\pi i}
\right)^{Dl/2} 
\Delta(\alpha)^{-D/2} 
\nonumber
\\&&
\times
\exp \biggl[ \sum^n_{i,j=1} 
\left\{
i \, k_i \cdot k_j Z_{ij} + 2 k_i \cdot \epsilon_j (\triangle_j Z_{ij})
-i \epsilon_i \cdot \epsilon_j (\triangle_i \triangle_j Z_{ij})
\right\}
\biggl]_{\mbox{linear in each $\epsilon$}}
\label{qedialpha2}
\end{eqnarray}
for an $l$-loop diagram with 
\begin{eqnarray}
\triangle_j Z_{ij} &=&
{\textstyle
\frac{Z_{ij'}-Z_{ij}}{2u_j'} + \frac{Z_{ij}-Z_{ij''}}{2u_j''} 
},
\\
\triangle_i \triangle_j Z_{ij} &=&
\frac{1}{4} \sum_{a,b}
\left( {\textstyle
\frac{\delta_{i'a}-\delta_{ia}}{u_i'} +
\frac{\delta_{ia}-\delta_{i''a}}{u_i''} 
}
\right)
\left( {\textstyle
\frac{\delta_{j'b}-\delta_{jb}}{u_j'} +
\frac{\delta_{jb}-\delta_{j''b}}{u_j''} 
}
\right)
Z_{ab}
\\
&=& \frac{1}{4u_i'u_j'} (Z_{i'j'}-Z_{ij'}-Z_{i'j}+Z_{ij}) + \ldots ~~ .
\end{eqnarray}
In the above expressions, $\Delta(\alpha)$ and $Z_{ij}$ are the same as
those appeared in eq.(\ref{ialpha}) for the $\phi^3$
diagram of the same topology, 
since we recover exactly eq.(\ref{ialpha1}) if we set all
$\epsilon_i =0$ and integrate out the dummy
vertices in eq.(\ref{qedialpha1}).
$Z_{ij'}$, etc.\ are defined similarly as in (\ref{defz}):
\begin{eqnarray}
g^{\mu\nu} Z_{ab}(\alpha) \equiv
- \frac{i}{4} 
\left< \! \left< \, (x_a-x_b)^\mu (x_a-x_b)^\nu \, \right> \! \right>,
\label{defz2}
\end{eqnarray}
but now $\left< \! \left< \ldots \right> \! \right>$ includes integral over
dummy vertices.

Remembering that $I(\alpha)$ is independent of $u_i'$ and $u_i''$,
we can take the limit $u_i',u_i'' \rightarrow +0$.
Due to the fact
\begin{eqnarray}
\lim_{u_i' \rightarrow 0} Z_{i'a} =
\lim_{u_i'' \rightarrow 0} Z_{i''a} =
Z_{ia} ,
\label{zbegin}
\end{eqnarray}
we can replace $\triangle_j Z_{ij}$ and
$\triangle_i \triangle_j Z_{ij}$ as
\begin{eqnarray}
\triangle_j Z_{ij} &=&
\frac{1}{2} 
\lim_{u_j', u_j''\rightarrow 0}
\left(
{\textstyle
\frac{\partial}{\partial u_j'} Z_{ij'}
- \frac{\partial}{\partial u_j''} Z_{ij''}
}
\right)
\label{triangle1}
\\
\triangle_i \triangle_j Z_{ij} &=&
\frac{1}{4}
\lim_{
\begin{array}{cc}
\scriptstyle
u_i',u_i'' \rightarrow 0\\
\scriptstyle
u_j',u_j'' \rightarrow 0
\end{array}
}
\left(
{\textstyle
\frac{\partial}{\partial u_i'} \frac{\partial}{\partial u_j'} Z_{i'j'} -
\frac{\partial}{\partial u_i'} \frac{\partial}{\partial u_j''} Z_{i'j''}
- 
\frac{\partial}{\partial u_i''} \frac{\partial}{\partial u_j'} Z_{i''j'} + 
\frac{\partial}{\partial u_i''} \frac{\partial}{\partial u_j''} Z_{i''j''} 
}
\right)
\label{triangle2}
\end{eqnarray}
At the same time, we can drop all diagonal terms ($i=j$) in
(\ref{qedialpha2}) using
\begin{eqnarray}
\lim_{u'_i \rightarrow +0} \,
\frac{\partial}{\partial u'_i} Z_{i'i}
\rule{0mm}{1.2cm}
= \lim_{u''_i \rightarrow +0} \,
\frac{\partial}{\partial u''_i} Z_{i''i}
\rule{0mm}{1.2cm}
= - \frac{1}{2}
\label{zend}
\end{eqnarray}
and noting that only the terms multi-linear in each $\epsilon_i$ should
be kept.
See Appendix B for proofs of eqs.(\ref{zbegin})-(\ref{zend}).
%ppppppppppppppppppppppppppppppppppppppppppppppppppppppppppppp
%
\begin{figure}
\begin{center}
\epsfile{file=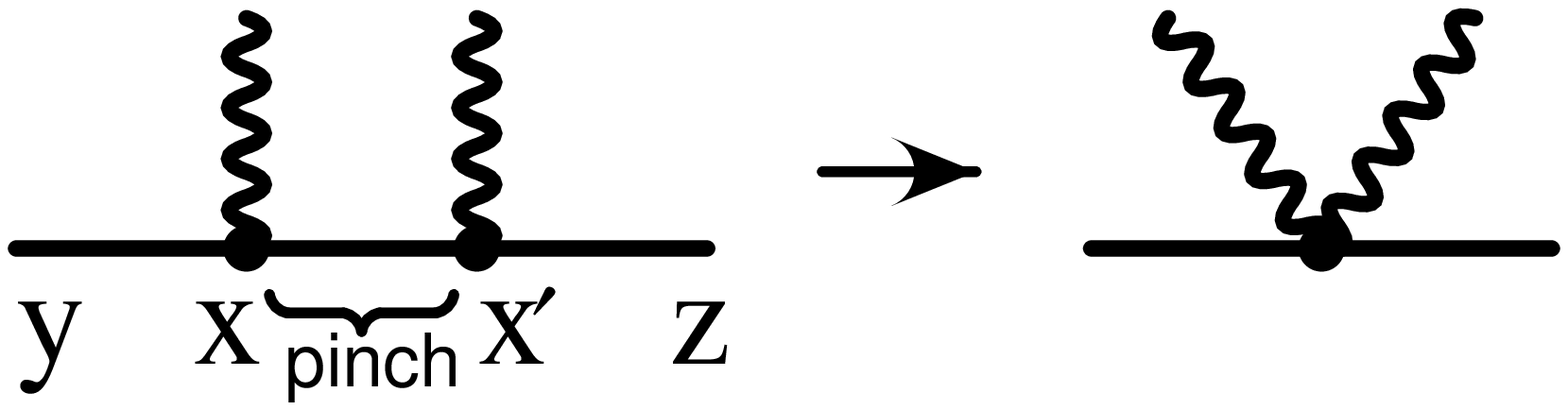,width=10cm}
\caption{The seagull vertex can be incorporated by pinching 
the propagator
between two adjacent three-point vertices with vertex factors 
$\epsilon^\mu e^{ik\cdot x}$ and $\epsilon'_\mu e^{ik'\cdot x}$.}
\end{center}
\end{figure}
%
%ppppppppppppppppppppppppppppppppppppppppppppppppppppppppppppp

So far we considered a diagram without seagull vertex.
The contribution of a seagull vertex can be incorporated through the
process known as
``pinching'' from the corresponding diagram without seagull vertex.
Any diagram containing a seagull vertex has the following factor
(see Fig.7):
\begin{eqnarray}
G_D(k,\epsilon) &\propto& i\Delta_F(y-x) \,
\epsilon^\mu e^{ik\cdot x} \, \epsilon'_\mu e^{ik'\cdot x} \,
i\Delta_F(x-z)
\\
&=&
\int dx' \, i\Delta_F(y-x) \,
\epsilon^\mu e^{ik\cdot x} \, \delta (x-x') \,
\epsilon'_\mu e^{ik'\cdot x'} \,
i\Delta_F(x'-z) .
\end{eqnarray}
The last line corresponds diagramatically to pinching
the propagator between the
two adjacent three-point vertices $x$ and $x'$; see Fig.7.
Noting that $\delta (x-x')$ is obtained by taking the 
$\alpha \rightarrow +0$ limit of the propagator in question
(see eq.(\ref{propk2})),
one can incorporate the contribution of a seagull vertex by replacing
\begin{eqnarray}
\epsilon_i \cdot \epsilon_j \, \triangle_i \triangle_j Z_{ij}
\rightarrow 2 \epsilon_i \cdot \epsilon_j \, \delta(\alpha_{ij}-0)
\label{pinch}
\end{eqnarray}
in eq.(\ref{qedialpha2}) of the diagram without seagull vertex,
where $\alpha_{ij}$ is the Feynman parameter
between the two adjacent three-point vertices $i$ and $j$.
If there are two or more seagull vertices in a diagram, one should 
pinch as
many propagators of the corresponding diagram without seagull vertex.

\subsection{Relation between General Expression and Feynman Parameter 
Formula}

Path-integral expression for $G_S(k,\epsilon)$ such as eq.(\ref{gi2}) 
can be obtained from the finite dimensional integral (\ref{qedialpha0}) 
by inserting infinitely many dummy vertices along scalar chains
using the associativity relation (\ref{assoc}).
The advantage of the path-integral expression lies in that it 
combines in a single expression sum of different diagrams that are
related to one another by sliding photon legs along the scalar
chains.
Different orderings of photon legs correspond to different orderings
of the proper time $t_i$'s of the vertices.

Once the ordering of $t_{i_l}$'s is fixed along 
the scalar chain $l$,
relations between $t_{i_l}$'s and Feynman parameters $\alpha_{i_l}$ are
given by:
\begin{itemize}
\item
For $l=$open, and 
$0 < t_1 < t_2 < \ldots < t_{n_l} < T_l$,
\begin{eqnarray}
\begin{array}{lcl}
t_1 &=& \alpha_1 \\
t_2-t_1 &=& \alpha_2  \\
&\vdots& \\
t_{n_l}-t_{n_l-1} &=& \alpha_{n_l}  \\
T_l-t_{n_l} &=& \alpha_{n_l+1} 
\end{array}
\label{ivtransf1}
\end{eqnarray}
\item
For $l=$closed, and 
$0 < t_1 < t_2 < \ldots < t_{n_l} < T_l$,
\begin{eqnarray}
\begin{array}{lcl}
t_1-t_{n_l}+T_l &=& \alpha_1 \\
t_2-t_1 &=& \alpha_2 \\
&\vdots& \\
t_{n_l}-t_{n_l-1} &=& \alpha_{n_l} 
\end{array}
\label{ivtransf2}
\end{eqnarray}
\end{itemize}
With these relations,
constituents of the general expression (\ref{generalexp}) and 
of the Feynman parameter formula (\ref{qedialpha2}) are
identified as follows:
\begin{eqnarray}
{\cal N} &=&
(2\pi)^D \delta \left( \sum^n_{i=1} k_i \right) \cdot
i^l \left( \frac{1}{4\pi i}
\right)^{Dl/2} 
\Delta^{-D/2} 
\label{ndelta}
\end{eqnarray}
and
\begin{eqnarray}
\begin{array}{rcr}
G_B^{ij} &=& - 2 \, Z_{ij} ,
\\
\rule{0mm}{5mm}
\partial_j G_B^{ij} &=& -2 \, \triangle_j Z_{ij} ,
\\
\rule{0mm}{5mm}
\partial_i \partial_j G_B^{ij} &=& -2 \, \triangle_i \triangle_j Z_{ij} .
\end{array}
\label{gbz}
\end{eqnarray}
We take the convention $G_B^{ii}=0$ in accord with the
zero-diagonal level scheme of $Z_{ab}$.
As $\cal N$ and $G_B^{ij}$'s are 
defined for a set of diagrams,
for a different ordering of $t_{i_l}$'s, $\Delta$ and 
$Z_{ij}$ of a different diagram should be taken on the right-hand-side.

It is more subtle how the contributions of seagull vertices are contained 
in the general expression (\ref{generalexp}).
They are contained in the $\partial_i \partial_j G_B^{ij}$ term when the
two vertices $t_i$ and $t_j$ come to the same point.
To see this, we consider the two-point function $G_B(\tau, \tau')$
defined in eq.(\ref{deftf}) when $\tau$ and $\tau'$ are arbitrary points 
along a same scalar chain.
One may, if necessary, identify it with 
$Z_{ab}$, 
where $x_a$ and $x_b$
are the dummy vertices inserted at the position of $\tau$ and $\tau'$,
respectively.
Due to eqs.(\ref{triangle1}), (\ref{triangle2}) and (\ref{gbz}),
one may express $G_B^{ij}$'s as
\begin{eqnarray}
G_B^{ij} &=& G_B(t_i,t_j)
\label{del0}
\\
\partial_j G_B^{ij} &=& 
\frac{1}{2} \biggl[ 
\lim_{\tau' \rightarrow t_j+0} 
+ \lim_{\tau' \rightarrow t_j-0} 
\biggl]
\frac{\partial}{\partial \tau'} \, G_B(t_i,\tau')
\label{delj}
\\
\partial_i \partial_j G_B^{ij} &=& 
\frac{1}{4} \biggl[ 
\lim_{\tau \rightarrow t_i+0} 
+ \lim_{\tau \rightarrow t_i-0} 
\biggl]
\biggl[ 
\lim_{\tau' \rightarrow t_j+0} 
+ \lim_{\tau' \rightarrow t_j-0} 
\biggl]
\frac{\partial}{\partial \tau} 
\frac{\partial}{\partial \tau'} \, G_B(\tau,\tau')
\label{delij}
\end{eqnarray}
for $i \neq j$, and we may omit all terms where $i=j$; see discussion
after eq.(\ref{triangle2}).
Then using the identity\footnote{
The corresponding identity
of $Z_{ab}$ is shown in Appendix B, eq.(\ref{proof4}).
}
\begin{eqnarray}
\lim_{\tau \rightarrow \tau'\pm 0} \,
\frac{\partial}{\partial \tau'}  \, G_B(\tau,\tau')
= \mp 1
\end{eqnarray}
which holds for any diagram,
it can be shown that
\begin{eqnarray}
{\hbox to 10pt{
\hbox to -3pt{$\displaystyle \int$} 
\raise-15pt\hbox to 7pt{$\scriptstyle t_j-u''$} 
\raise18pt\hbox{$\scriptstyle t_j+u'$}
}} ~
dt_i \, \, \partial_i \partial_j G_B^{ij}
= -2 + 
\biggl(
{\hbox to 10pt{
\hbox to -3pt{$\displaystyle \int$} 
\raise-15pt\hbox to 7pt{$\scriptstyle t_j+0$} 
\raise18pt\hbox{$\scriptstyle t_j+u'$}
}}
~dt_i +
{\hbox to 10pt{
\hbox to -3pt{$\displaystyle \int$} 
\raise-15pt\hbox to 7pt{$\scriptstyle t_j-u''$} 
\raise18pt\hbox{$\scriptstyle t_j-0$}
}}
~dt_i~
\biggl) \, 
\partial_i \partial_j G_B^{ij}
~~~~~(u',u''>0) .
\end{eqnarray}
Thus, we see $\delta$-function contribution as
\begin{eqnarray}
\partial_i \partial_j G_B^{ij} \sim -2 \delta(t_i-t_j)
~~~~~\mbox{for}~~
t_j-0 < t_i < t_j+0 ,
\end{eqnarray}
so that the contributions of seagull vertices are included as in
eq.(\ref{pinch}).
(The factor 2 is accounted for by the interchange of $i$ and $j$.)
It is interesting how gauge symmetry takes advantage of the property of
$G_B(\tau,\tau')$ which is an intrinsic quantity to any diagram.

Finally we comment on the integral variables 
of the two formulas (\ref{generalexp}) and
(\ref{gd}). 
Note that along a closed scalar 
chain we have one more time variables to integrate over
($t_1, \ldots , t_{n_l}, T_l$)
than the corresponding Feynman parameters.
In fact, one proper time variable can be integrated trivially;
after the first $n_l-1$ integrals over $t_{i_l}$'s, there remains no
dependence on $t_{n_l}$\footnote{
Any function of the form
\begin{eqnarray}
f(t_{n_l}) = \int^{T_l}_0 dt_{n_l-1} \cdots \int^{T_l}_0 dt_1 \,
F( G_B^{ij}, {\cal N} )
~~~~~(l:\mbox{closed chain})
\end{eqnarray}
is invariant under translation 
$t_{n_l} \rightarrow t_{n_l} + c$
since $G_B^{ij}$ and ${\cal N}$ are periodic functions of $t_{i_l}$'s
and depend only on $t_{i_l}-t_{j_l}$; 
see eqs.(\ref{ndelta}) and (\ref{gbz}).
This means $f'(t)=0$ so that $f(t)$ is independent of $t$.
},
so the last integral just gives a factor of $T_l$, which compensates
$T_l^{-1}$ in the integral measure (\ref{imeasure}).

\subsection{Decomposition of $G_B$ and $\cal N$}
\clfn

Up to now we dealt with $G_B(\tau,\tau')$ and $\cal N$ for a general set
of diagrams.
We show that these quantities can be decomposed and written in terms of 
those for the basic
sets of diagrams, namely, $G_B(\tau,\tau')$ and $\cal N$ for an open
scalar chain and for a closed scalar chain; see Fig.8.
%ppppppppppppppppppppppppppppppppppppppppppppppppppppppppppppp
%
\begin{figure}
\hspace*{4cm}
\epsfile{file=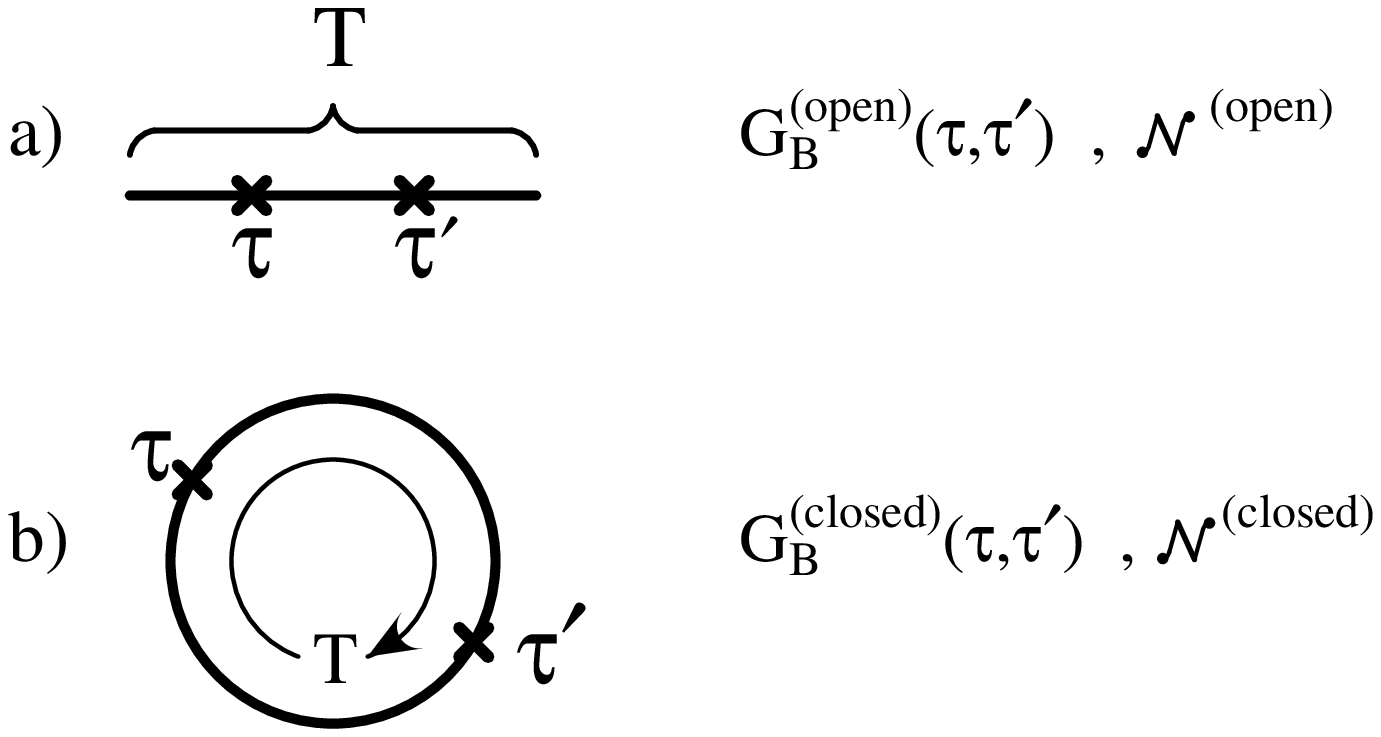,height=5cm}
\caption{The basic diagrams: a) an open scalar chain, and 
b) a closed scalar chain.
Two-point function for an arbitrary set of diagrams 
can be decomposed and written
in terms of $G_B^{(open)}$ and $G_B^{(closed)}$.}
\end{figure}
%
%ppppppppppppppppppppppppppppppppppppppppppppppppppppppppppppp
%ppppppppppppppppppppppppppppppppppppppppppppppppppppppppppppp
%
\begin{figure}
\begin{center}
\epsfile{file=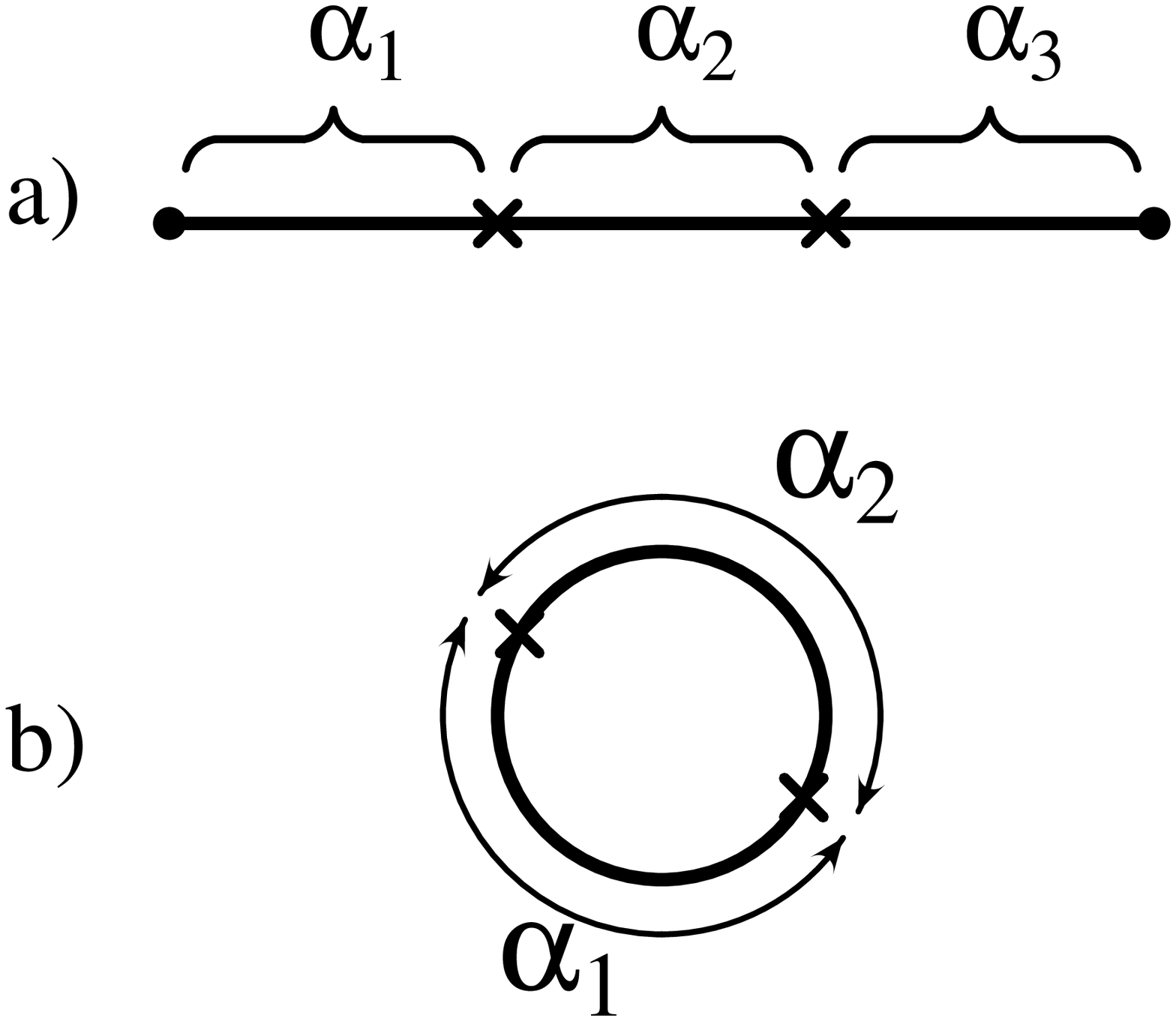,width=7cm}
\caption{The basic diagrams corresponding to Fig.8 
but parametrized by Feynman
parameters.}
\end{center}
\end{figure}
%
%ppppppppppppppppppppppppppppppppppppppppppppppppppppppppppppp

Let us first find the explicit forms of these basic
$G_B(\tau,\tau')$ and $\cal N$. 
They are obtained from $Z_{ij}$ and $\Delta(\alpha)$ for
the corresponding diagrams (Fig.9).
According to the calculation method described in Appendix B,
one obtains for these diagrams
\begin{eqnarray}
&&Z_{12}^{(open)} = -\frac{1}{2} \, \alpha_2, 
~~~~~ ~~~~~ ~~~~~
\Delta^{(open)} = 1 ,
\\
&&Z_{12}^{(closed)} 
= -\frac{1}{2} \, \frac{\alpha_1\alpha_2}{\alpha_1+\alpha_2},
~~~~~~
\Delta^{(closed)} = \alpha_1+\alpha_2 .
\end{eqnarray}
It follows
\begin{eqnarray}
&&G_B^{(open)}(\tau,\tau') = | \tau - \tau' |,
~~~~~~~~~~~~~~~~~~~~~~~~
\Delta^{(open)} = 1 ,
\\
&&G_B^{(closed)}(\tau,\tau') = | \tau - \tau' |
- \frac{(\tau-\tau')^2}{T} ,
~~~~~~
\Delta^{(closed)} = T ,
\end{eqnarray}
where the normalization factor $\cal N$ is given by
eq.(\ref{ndelta}).

We deal with finite dimensional integral, and start from the defining
equation of $Z_{ab}$ and $\Delta$ for a diagram $D$:
\begin{eqnarray}
I &=& \int [dx_a] \, \exp \biggl[
-\frac{i}{4} \sum_{a,b} x_a \cdot x_b A_{ab} + i \sum_a J_a \cdot x_a 
\biggl]
\label{defzdelta1}
\\
&=&
(2\pi)^D \, \delta \biggl( \sum_a J_a \biggl)
\cdot
i^l \left( \frac{1}{4\pi i}
\right)^{Dl/2} 
\Delta^{-D/2} \,
\exp \biggl[ i \sum_{a,b} J_a \cdot J_b Z_{ab} \biggl] .
\label{defzdelta2}
\end{eqnarray}
We would like to know how the above expression changes when the vertices 
$i$ and $j$ in $D$ are connected by a propagator whose Feynman parameter 
is $\alpha$.
(The diagram thus obtained is denoted as $D'$.)
This is achieved if we multiply the integrand in (\ref{defzdelta1})
by
\begin{eqnarray}
K(x_i-x_j;\alpha) 
&=& i \left( \frac{1}{4\pi i\alpha} \right)^{D/2}
\exp \left[ -\frac{i}{4\alpha}(x_i-x_j)^2 \right] 
\end{eqnarray}
before integration over $[dx_a]$.
But it is an equivalent manipulation if we shift
\begin{eqnarray}
J_a \rightarrow J_a + p \,( \delta_{ai} -\delta_{aj} ) ,
\end{eqnarray}
multiply by $\exp ( i\alpha p^2 )$,
and then integrate over $p$; see eq.(\ref{nrp1}).
Applying this manipulation to (\ref{defzdelta2}), one obtains
\begin{eqnarray}
I~ \rightarrow ~ I' &=&
(2\pi)^D \, \delta \biggl( \sum_a J_a \biggl)
\cdot
i^{l+1} \left( \frac{1}{4\pi i}
\right)^{D(l+1)/2} 
\left[ \Delta \cdot (\alpha - 2 \, Z_{ij}) \right]^{-D/2} 
\\ && ~~~
\times
\exp \biggl[ i \sum_{a,b} J_a \cdot J_b 
\biggl\{ Z_{ab} +
\frac{ ( Z_{ia}-Z_{ja}-Z_{ib}+Z_{jb} )^2 }
{ 2 \, ( \alpha - 2 \, Z_{ij} ) }
\biggl\}
\biggl] .
\end{eqnarray}
This expression defines $\Delta$ and $Z_{ab}$ for $D'$, and
correspondingly we find the following rule\footnote{
Eq.(\ref{decgb1}) differs slightly
from the expression obtained by Schmidt and Schubert\cite{ss}
since they do not take the zero-diagonal level scheme.
The difference is accounted for by the transformation
(\ref{transfz}).
}
for obtaining
$\cal N$ and $G_B$ for the diagram $D'$:
\begin{eqnarray}
\Delta' &=& 
{\Delta} \cdot 
\left( \alpha + \, {G}_B(t_i,t_j) \right) ,
\label{decdel1}
\\
G_B'(\tau,\tau') &=& {G}_B(\tau,\tau') -
\frac{
\left( {G}_B(\tau,t_i) - {G}_B(\tau,t_j)
- {G}_B(\tau',t_i) + {G}_B(\tau',t_j) \right)^2
}{ 4 \left( \alpha + \, {G}_B(t_i,t_j) \right) } .
\label{decgb1}
\end{eqnarray}

Next we consider the case where two diagrams 
$D_1 (\ni i)$ and $D_2 (\ni j)$ are sewn
together by a propagator $(ij)$.
In this case, we shift
\begin{eqnarray}
J^{(1)}_a \rightarrow J^{(1)}_a + p \, \delta_{ia} ,
~~~~~
J^{(2)}_a \rightarrow J^{(2)}_a - p \, \delta_{ja}
\end{eqnarray}
in $I^{(1)}$ and $I^{(2)}$, respectively, multiply by
$\exp ( i\alpha p^2 )$,
and then integrate over $p$.
It is straightforward to find the following rule:
\begin{eqnarray}
\Delta' &=& \Delta^{(1)} \cdot \Delta^{(2)} ,
\label{decdel2}
\\
G_B'(\tau,\tau') &=& \left\{
\begin{array}{ll}
\alpha + G_B^{(1)}(\tau,t_i) + G_B^{(2)}(\tau',t_j)
~~~
& \tau \in D_1,~ \tau' \in D_2 \\
\rule{0mm}{6mm}
G_B^{(1)}(\tau,\tau')
& \tau,\tau' \in D_1 \\
\rule{0mm}{6mm}
G_B^{(2)}(\tau,\tau')
& \tau,\tau' \in D_2 
\end{array}
\right. .
\label{dec}
\end{eqnarray}

Any set $S$ of diagrams can be constructed by connecting scalar chains with
photon propagators.
Then one may express $G_B$ ($\cal N$) for $S$ in terms of 
$G_B^{(open)}$ (${\cal N}^{(open)}$) and $G_B^{(closed)}$
(${\cal N}^{(closed)}$)
either by using the above rules recursively, or,
by applying similar
manipulation for multiple photon propagator insertions at once.

Now we find an important property of two-point functions 
$\partial_j G_B^{ij}$ and $\partial_i \partial_j G_B^{ij}$.
Writing $G_B(\tau,\tau')$ for an arbitrary set of diagrams in terms of
the basic elements, we notice that
$\partial_i$ ($\partial_j$) can
be replaced by $\partial /\partial t_i$ ($\partial /\partial t_j$)
if the vertex $i$ ($j$) is external\cite{lam} or if
the diagram is one-particle-reducible with respect to the photon
propagator connected to the vertex $i$ ($j$).
(cf.\ eqs.(\ref{delj}) and (\ref{delij}).)

\section{Integration By Parts}

\cleqn

Now we are in place to explain 
the integration by parts technique, 
first introduced to field theoretical
calculation by Bern and Kosower, which 
enables non-trivial reshuffling of
various terms in eq.(\ref{generalexp}) {\it
before} integrating over $\alpha_r$, $t_{i_l}$, and $T_l$.
This technique reduces the number of
independent terms, and consequently reduces
the labor in
the evaluation of integrals.

\subsection{Example}

%ppppppppppppppppppppppppppppppppppppppppppppppppppppppppppppp
%
\begin{figure}
\begin{center}
\epsfile{file=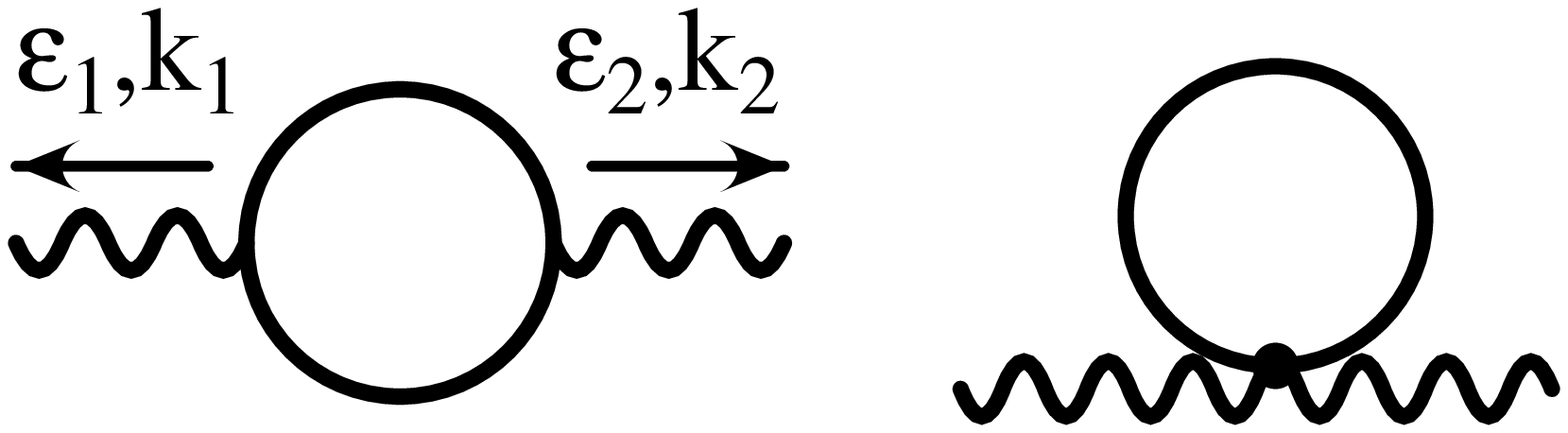,width=10cm}
\caption{The one-loop diagrams contributing to the photon vacuum polarization.}
\end{center}
\end{figure}
%
%ppppppppppppppppppppppppppppppppppppppppppppppppppppppppppppp

Consider a simplest example \cite{strassler1}.
According to eq.(\ref{generalexp}) and the manipulation 1)-4),
the photon vacuum polarization at one-loop (Fig.10) is given by
\begin{eqnarray}
G_S &=& (2\pi)^D \delta (k_1+k_2) \cdot
(ie)^2 \cdot i 
\left( \frac{1}{4\pi i} \right)^{D/2} 
\int^\infty_0 \frac{dT}{T}
\int^T_0 dt_1 \, dt_2
\nonumber \\
&& ~~~
\times T^{-D/2} \,  e^{-ik_1\cdot k_2 G_B^{12}} \,
( k_1\cdot \epsilon_2 \, k_2 \cdot \epsilon_1 \, 
\partial_1 G_B^{12}\, \partial_2
G_B^{12} 
+ i \epsilon_1 \cdot \epsilon_2 \, \partial_1 \partial_2 G_B^{12} ),
\end{eqnarray}
where we used $\Delta = T$.
Note that $\partial_1$ ($\partial_2$) can be identified with
$\partial /\partial t_1$ ($\partial /\partial t_2$) 
since vertices 1 and 2 are external vertices.
We integrate by parts the second term with respect to $t_1$.
The surface term vanishes due to the periodicity of $G_B^{ij}$.
Thus,
\begin{eqnarray}
G_S &=& - (2\pi)^D \delta (k_1+k_2) \cdot
ie^2 \cdot \left( \frac{1}{4\pi i} \right)^{D/2} 
(  k_1\cdot \epsilon_2 \, k_2 \cdot \epsilon_1 \, -
\epsilon_1 \cdot \epsilon_2 \, k_1 \cdot k_2 )
\nonumber \\
&& ~~~
\times \int^\infty_0 dT \, T^{-1-D/2}
\int^T_0 dt_1 \, dt_2 \,
e^{-ik_1\cdot k_2 G_B^{12}} \,
\partial_1 G_B^{12}\, \partial_2
G_B^{12} ,
\label{partialint}
\end{eqnarray}
and we find $G_S$ is gauge-invariant {\it before} integration over 
$t_1$, $t_2$ and $T$.
Note that the number of independent terms reduced from two to one.

To see the relation between gauge transformation and the integration by
parts technique, we remember
\begin{eqnarray}
G_S \propto
\left<
\int^T_0 dt_1 \, \epsilon_1 \cdot \dot{x}(t_1) \, e^{ik_1\cdot x(t_1)}
\times
\int^T_0 dt_2 \, \epsilon_2 \cdot \dot{x}(t_2) \, e^{ik_2\cdot x(t_2)}
\right>,
\label{pathgs}
\end{eqnarray}
where $\left< \ldots \right>$ denotes the path-integral average.
Gauge transformation of photon 1 is achieved by replacing 
$\epsilon_1$ by $k_1$.
Then the vertex operator changes as
\begin{eqnarray}
\epsilon_1 \cdot \dot{x}(t_1) \, e^{ik_1\cdot x(t_1)}
\rightarrow k_1 \cdot \dot{x}(t_1) \, e^{ik_1\cdot x(t_1)}
= -i \frac{d}{dt_1} e^{ik_1\cdot x(t_1)},
\end{eqnarray}
and
\begin{eqnarray}
\delta G_S &\propto&
\left<
\int^T_0 dt_1 \, \frac{d}{dt_1} e^{ik_1\cdot x(t_1)}
\times
\int^T_0 dt_2 \, \epsilon_2 \cdot \dot{x}(t_2) \, e^{ik_2\cdot x(t_2)}
\right>
\nonumber
\\
&=&
\int^T_0 dt_1 \, dt_2 \, \frac{\partial}{\partial t_1}
\left<
e^{ik_1\cdot x(t_1)}
\,
\epsilon_2 \cdot \dot{x}(t_2) \, e^{ik_2\cdot x(t_2)}
\right>
\nonumber \\
&=&
\int^T_0 dt_1 \, dt_2 \, \frac{\partial}{\partial t_1}
\, \left(- k_1 \cdot \epsilon_2 \, \partial_2 G_B^{12} \, 
e^{-ik_1\cdot k_2 G_B^{12}} \right) .
\end{eqnarray}
Gauge transform of the integrand is given by total derivative, so 
$G_S$ is
obviously gauge-invariant whereas the integrand itself
is not.
We may add, however, to the integrand of $G_S$ 
in eq.(\ref{pathgs}) a term which 
transforms equally but in opposite sign under the replacement
$\epsilon_1 \rightarrow k_1$:
\begin{eqnarray}
\frac{\partial}{\partial t_1} 
\left( \epsilon_1 \cdot \epsilon_2 \,
\partial_2 G_B^{12} \, 
e^{-ik_1\cdot k_2 G_B^{12}} \right) .
\end{eqnarray}
Being total derivative, addition of this term does not alter $G_S$.
Now the integrand itself is gauge-invariant, and
the above term is exactly the surface term of the partial integration 
in eq.(\ref{partialint}).

\subsection{External Photon}

We now show a general prescription for integration by parts with
respect to the external gauge vertices.

First, if the external photons are on-shell and for a fixed helicity
states, one can use spinor helicity 
technique\cite{sht1,sht2} 
to reduce the number of dot products in the exponent of 
the general expression (\ref{generalexp}).
On the other hand, if the external photons are off-shell, one can replace 
each polarization vector as
\begin{eqnarray}
\epsilon_i^\mu \rightarrow {\epsilon'_i}^\mu = \epsilon_i^\mu
- \frac{\epsilon_i \cdot k_a}{k_i \cdot k_a} \, k_i^\mu
= (\epsilon_i^\mu k_i^\nu - k_i^\mu \epsilon_i^\nu) \, k_{a\nu} \,
\frac{1}{k_i \cdot k_a} .
\label{offshellspinhel}
\end{eqnarray}
The amplitude is invariant under this replacement, and also
the resulting expression is manifestly gauge-invariant before
integration over proper time variables.
One may choose any $k_a$ for each polarization vector $\epsilon_i$.
Since $k_a \cdot \epsilon'_i =0$,
appropriate choices of $k_a$'s for all $i$'s will reduce the number of
terms in the exponent.

After reducing the terms in the exponent, and after manipulation 
1)-4) above eq.(\ref{replpolv}), one integrates by parts with respect 
to the proper time of external vertices to reduce the number of
independent terms in the integrand.
In this procedure, one may omit surface terms for a closed scalar
chain since the surface terms cancel with each other due to the 
periodicity of $G_B$.
Also for an open scalar chain, surface terms can be neglected if one is
interested in the $S$-matrix element, since each surface 
term cancel the
propagator pole of the external scalars in the
unamputated Green function; see Fig.11.
%ppppppppppppppppppppppppppppppppppppppppppppppppppppppppppppp
%
\begin{figure}
\begin{center}
\epsfile{file=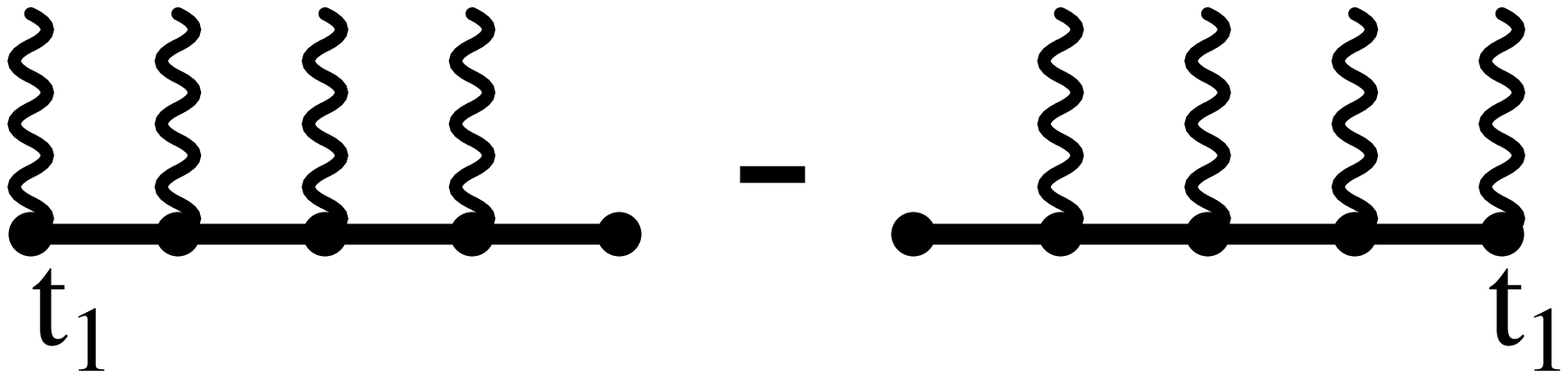,width=10cm}
\caption{The surface terms originating from the 
gauge transformation of an
external photon along an open chain. 
(Some of) The propagator poles of external scalars 
get cancelled, so these surface terms 
do not contribute to the $S$-matrix element.}
\end{center}
\end{figure}
%
%ppppppppppppppppppppppppppppppppppppppppppppppppppppppppppppp

\subsection{Internal Photon}

One may also apply integration by parts technique to the internal gauge
vertices.\cite{ss}
Using the decomposition rule derived in the previous section, one can
write $\Delta$, 
$\partial_j G_B$ and $\partial_i \partial_j G_B$ using
$G_B^{(open)}$, $G_B^{(closed)}$, and their
derivatives.
One can always integrate by parts to eliminate all second derivatives.
This corresponds to simplifying the expression using gauge
transformation of the internal vertices.

There is one exception for this procedure.  
The integration by parts with respect to any of the internal vertices 
whose the other end of the photon propagator is on a same 
{\it open} scalar chain
does not lead to simplification.
The surface terms of such partial integration still comprise
the poles of external scalars as seen in Fig.12.
Thus, one cannot omit the surface terms in this case.

%ppppppppppppppppppppppppppppppppppppppppppppppppppppppppppppp
%
\begin{figure}
\begin{center}
\epsfile{file=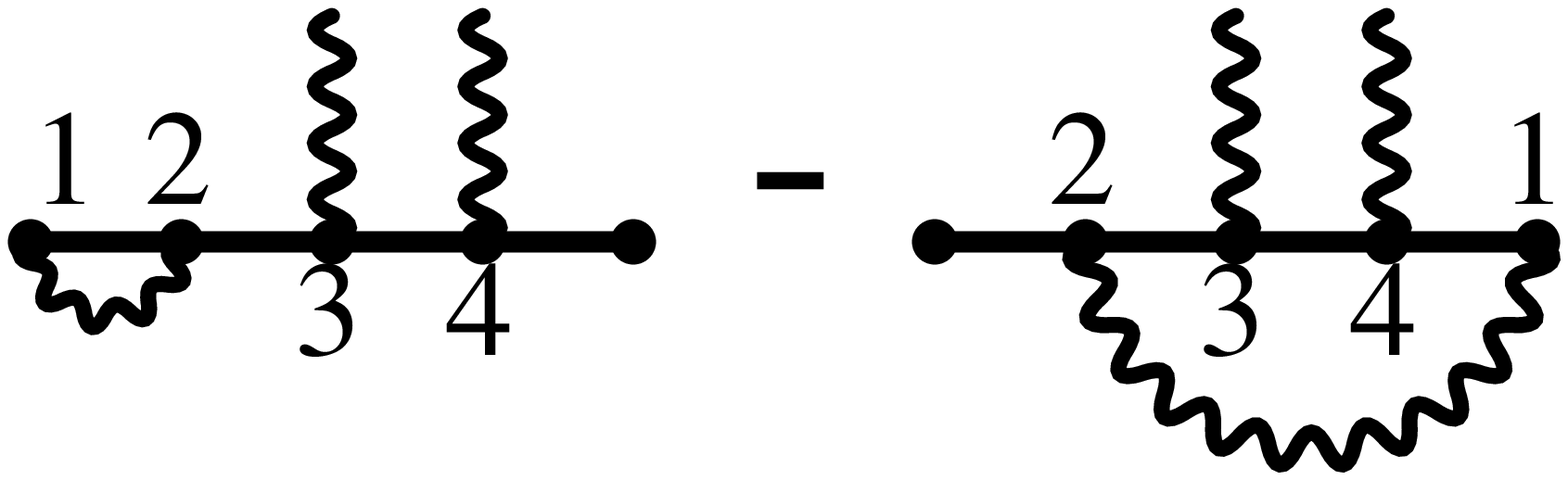,width=10cm}
\caption{The surface terms originating from the gauge
transformation of an internal photon whose both ends are attached to a
same open scalar chain.
(Some of) The surface terms cannot be omitted
since they still
contain the propagator poles of external scalars.}
\end{center}
\end{figure}
%
%ppppppppppppppppppppppppppppppppppppppppppppppppppppppppppppp

\section{Covariant Gauge for Internal Photons}

\cleqn

%ppppppppppppppppppppppppppppppppppppppppppppppppppppppppppppp
%
\begin{figure}
\begin{center}
\epsfile{file=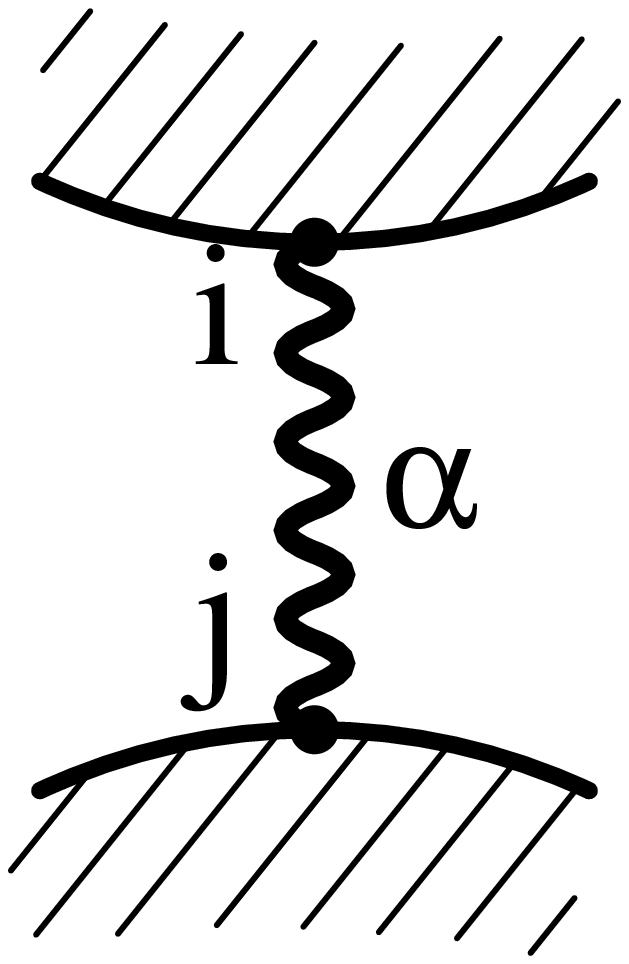,width=4cm}
\caption{The covariant gauge photon propagator 
whose Feynman parameter is $\alpha$. }
\end{center}
\end{figure}
%
%ppppppppppppppppppppppppppppppppppppppppppppppppppppppppppppp

From a field theoretical point of view it is interesting to know how
the general expression changes if one used covariant gauge for internal
photon propagators instead of Feynman gauge.
Let $i$ and $j$ be the vertices at 
the both ends of photon propagator whose Feynman parameter is $\alpha$;
see Fig.13.
In momentum space it can be written as
\begin{eqnarray}
\frac{-i}{p^2+i\epsilon} \, \biggl[ \,
g_{\mu\nu} - (1-\xi)\frac{p_\mu p_\nu}{p^2} \, \biggl] .
\end{eqnarray}
The $g_{\mu\nu}$ part is the Feynman gauge propagator, and appears in
the path-integral formalism as
\begin{eqnarray}
\dot{x}_i^\mu \, \dot{x}_j^\nu \, \,  g_{\mu\nu}
\exp \biggl[ -\frac{i}{4\alpha} (x_i-x_j)^2 \biggl] 
\end{eqnarray}
with $x_i \equiv x(t_i)$ and $x_j \equiv x(t_j)$.
Meanwhile, $p_\mu p_\nu$ part can be written as
\begin{eqnarray}
\dot{x}_i^\mu \, \dot{x}_j^\nu \, \,
i \alpha \frac{\partial}{\partial x_i^\mu} 
\frac{\partial}{\partial x_j^\nu} \,
\exp \biggl[ -\frac{i}{4\alpha} (x_i-x_j)^2 \biggl] 
=
i \alpha \,
\frac{\partial}{\partial t_i}\frac{\partial}{\partial t_j}
\exp \biggl[ -\frac{i}{4\alpha} (x_i-x_j)^2 \biggl] ,
\end{eqnarray}
where we used
\begin{eqnarray}
i \int^\infty_0 d\alpha \, \alpha 
\, \int \frac{d^Dp}{(2\pi)^D} \, 
\frac{\partial}{\partial x^\mu}\frac{\partial}{\partial y^\nu}
\, e^{ip \cdot (x-y) + i \alpha p^2}
= -i  \int \frac{d^Dp}{(2\pi)^D} \, 
\frac{p_\mu p_\nu}{p^4} \, e^{ip \cdot (x-y)} ,
\end{eqnarray}
cf.\ eq.(\ref{fpfp2}).
Therefore, we obtain the $p_\mu p_\nu$ part of photon $(ij)$ 
by operating
\begin{eqnarray}
(1-\xi) \, i\alpha 
\frac{\partial}{\partial t_i}\frac{\partial}{\partial t_j}
\end{eqnarray} 
to the integrand of eq.(\ref{generalexp})
after setting 
$\epsilon_i = \epsilon_j = 0$.
Again this is given by total derivative,
so changing gauge parameter $\xi$ can be regarded as a kind of gauge 
transformation.

From this we see that if one calculates a set of diagrams in different
values of $\xi$,
the difference of results is proportional to the surface term on 
each scalar chain.
In particular, a set of diagrams without external scalars is
independent of $\xi$ (if expressed in terms of bare coupling 
and bare gauge
parameter) since $G_B(\tau,\tau')$ is 
periodic function on each closed
scalar chain.

\section{Rule}
\cleqn

Let us summarize the Bern-Kosower-type rule for calculating 
a set of diagrams in Scalar QED
({\it amputated} with respected to external photons
and {\it unamputated} with respect to external scalars).
The gauge-invariant sub-Green function for a set $S$ with $2n_s$
external scalars at ${\cal O}(e^n)$ and for $l$ loop
is given by
\begin{eqnarray}
G_S(k,\epsilon) &=& 
(2\pi)^D \delta(\sum k_i) \cdot i^l \left( \frac{1}{4\pi i}
\right)^{Dl/2} 
(ie)^n \, C \,
\nonumber \\
&& \times
\int^\infty_0 \prod_r d\alpha_r \, \prod_{chain \, l}
\left(
\int^\infty_0 [dT_l]\, e^{-i(m^2-i0)T_l} 
\int^{T_l}_0 \prod_{i_l} dt_{i_l}
\right)
\, {\cal K}_{red},
\label{rule1}
\end{eqnarray}
where $C$ is the combinatorial factor,
$\alpha_r$ denotes the Feynman
parameter of the $r$-th photon propagator.
The chain $l$ represents
open or closed scalar chain, 
and the integral measure for its length
$T_l$ is 
\begin{eqnarray}
[dT_l] = \left\{
\begin{array}{cl}
dT_l & \mbox{for $l=$open}\\
\rule{0mm}{4mm} {dT_l}/{T_l} & \mbox{for $l=$closed}
\end{array} 
\right. .
\end{eqnarray}
$i_l$ represents photon vertex on the chain $l$.

The so-called reduced generating kinematical factor ${\cal
K}_{red}$ is obtained from the generating kinematical
factor
\begin{eqnarray}
{\cal K} &=&
\Delta^{-D/2} \cdot \exp \biggl[ \, \frac{1}{2}
\sum_{i\neq j}^{n+2n_s} \biggl\{
- i k_i \cdot k_j G_B^{ij}
- 2 k_i \cdot \epsilon_j \partial_j G_B^{ij}
+ i \epsilon_i \cdot \epsilon_j \partial_i \partial_j G_B^{ij}
\biggl\}
\biggl] 
\label{gkf}
\end{eqnarray}
after the following manipulation.
\begin{enumerate}
\renewcommand{\labelenumi}{\arabic{enumi})}
\item 
If the vertex $i$ is internal, we set corresponding $k_i=0$. 
\item
If the vertex $i$ is an endpoint of an open scalar chain,
we set corresponding $\epsilon_i=0$.
\item
If the external photons are on-shell and for a fixed helicity states,
use spinor helicity technique to reduce the number of dot products in
the exponent; if the external photons are off-shell use replacement
(\ref{offshellspinhel}) to reduce the number of dot products
(written in terms of $\epsilon'_i$'s).
\item
Only the terms multi-linear in each remaining polarization vector
are kept.
\item
We replace the polarization vectors at both ends of every photon
propagator $r$ as
\begin{eqnarray}
\epsilon_{i_r}^\mu \epsilon_{j_r}^\nu \rightarrow -g^{\mu \nu} .
\end{eqnarray} 
Again some of the Lorentz contractions vanish.
\end{enumerate}

Then integrate by parts with respect to the proper-times of external
vertices.
Also, integrate by parts with respect to the proper-times of internal
vertices after writing $\Delta$, 
$\partial_j G_B$ and $\partial_i \partial_j G_B$ 
in terms of $G_B^{(open)}$, $G_B^{(closed)}$, and their
derivatives. 
(Use decomposition rules (\ref{decdel1}), (\ref{decgb1}),
(\ref{decdel2}) and (\ref{dec}), and also eqs.(\ref{del0})-(\ref{delij}) 
for this purpose.)
Surface terms can be omitted except for the special 
case described in subsection 4.c.
The partial integrations generally reduce the number of independent
terms.

In order to integrate over $\alpha_r$, $t_i$, and $T_l$,
it is sometimes convenient to transform the variables
to the conventional Feynman 
parameter at this stage using relations (\ref{ivtransf1}) and
(\ref{ivtransf2}).

\section{Operator Insertion}
\cleqn

So far we have considered sets of diagrams containing only gauge
interactions. 
In practical calculations, however,
one will need to calculate diagrams containing 
both gauge interactions and other interactions, or more generally,
operator insertions to the sets of diagrams considered above.
We show in two examples how to calculate such diagrams.
The idea is to replace any operator ${\cal O}(\phi)$ 
by the functional derivatives
$\delta/\delta J(x)$ and $\delta/\delta J^*(x)$.

%ppppppppppppppppppppppppppppppppppppppppppppppppppppppppppppp
%
\begin{figure}
\begin{center}
\epsfile{file=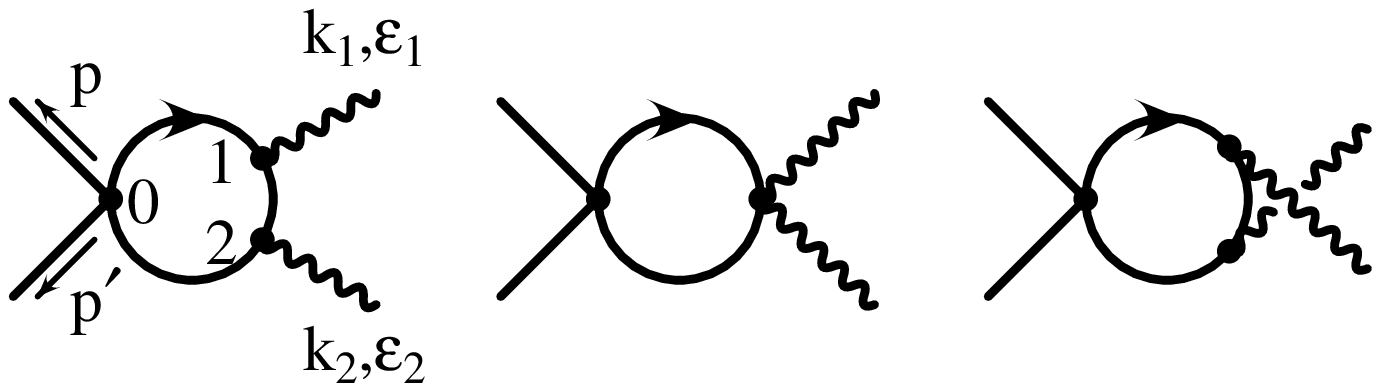,width=13cm}
\caption{A set of one-loop diagrams containing a 
$\phi^4$-operator insertion.}
\end{center}
\end{figure}
%
%ppppppppppppppppppppppppppppppppppppppppppppppppppppppppppppp

Let us see how to calculate the set of diagrams in Fig.14 contributing to 
the Green function with a $|\phi|^4$ operator insertion:
\begin{eqnarray}
&&
\left.
\int \! {\cal D}\phi \, {\cal D}Q_\mu \,
\int dz \, \frac{i\lambda}{4} |\phi(z)|^4 \,
\exp \, {i \! \int dx \,
[ {\cal L} + {\cal L}_{gf}
+ J^*\phi + J\phi^* + j^\mu Q_\mu ]
} \,
\right|_{j_\mu \rightarrow - \Box A_\mu} 
\\
&& = \frac{i\lambda}{4} \int dz \,
\left( \frac{\delta}{\delta J(z)} \right)^2
\left( \frac{\delta}{\delta J^*(z)} \right)^2
e^{W(J,J^* \! \! ,A_\mu)} .
\label{opgenerfn}
\end{eqnarray}
Following similar steps as in 
eqs.(\ref{defgreen})-(\ref{egfinal}), we find
\begin{eqnarray}
G(k,\epsilon) &=&
(2\pi)^D \delta (\sum k_i ) \cdot i 
\left( \frac{1}{4\pi i} \right)^{D/2} \! \! \cdot
(i\lambda) (ie)^2 \int^\infty_0 dT \, e^{-im^2T} \int^T_0 dt_1 dt_2
\nonumber
\\
&& \times
\Delta \, \exp \biggl[ \, \frac{1}{2}
\sum_{i \neq j} \biggl\{
- i k_i \cdot k_j G_B^{ij}
- 2 k_i \cdot \epsilon_j \partial_j G_B^{ij}
+ i \epsilon_i \cdot \epsilon_j \partial_i \partial_j G_B^{ij}
\biggl\}
\biggl] , 
\end{eqnarray}
where $k_0 = p+p'$ and $\epsilon_0=0$.
The two-point function $G_B(\tau,\tau')$ 
is obtained using the decomposition rule described in
subsection 3.d with a little modification.
Namely, we can compute $G_B$ by connecting both ends of an
open scalar chain with a
dummy photon propagator, and then pinching the photon
propagator by setting its
Feynman parameter as $\alpha \rightarrow 0$; see Fig.15 and
eq.(\ref{propk2}). 
Therefore, we find using (\ref{decgb1})
\begin{eqnarray}
G_B(\tau,\tau') &=& |\tau-\tau'| -
\frac{[ \tau - (T-\tau)-\tau'+(T-\tau')]^2}{4T}
\nonumber
\\
&=& |\tau-\tau'| - \frac{(\tau-\tau')^2}{T}
\end{eqnarray}
and
\begin{eqnarray}
\Delta = T .
\end{eqnarray}
The above two-point function coincides with $G_B^{(closed)}$, which
is a reasonable result.
Note, however, that the integral measure $dT$ differs from that of a
closed scalar
chain since the zeroth vertex is not that of gauge interaction.
Compare the discussion in the last paragraph in subsection 3.c.
%ppppppppppppppppppppppppppppppppppppppppppppppppppppppppppppp
%
\begin{figure}
\begin{center}
\epsfile{file=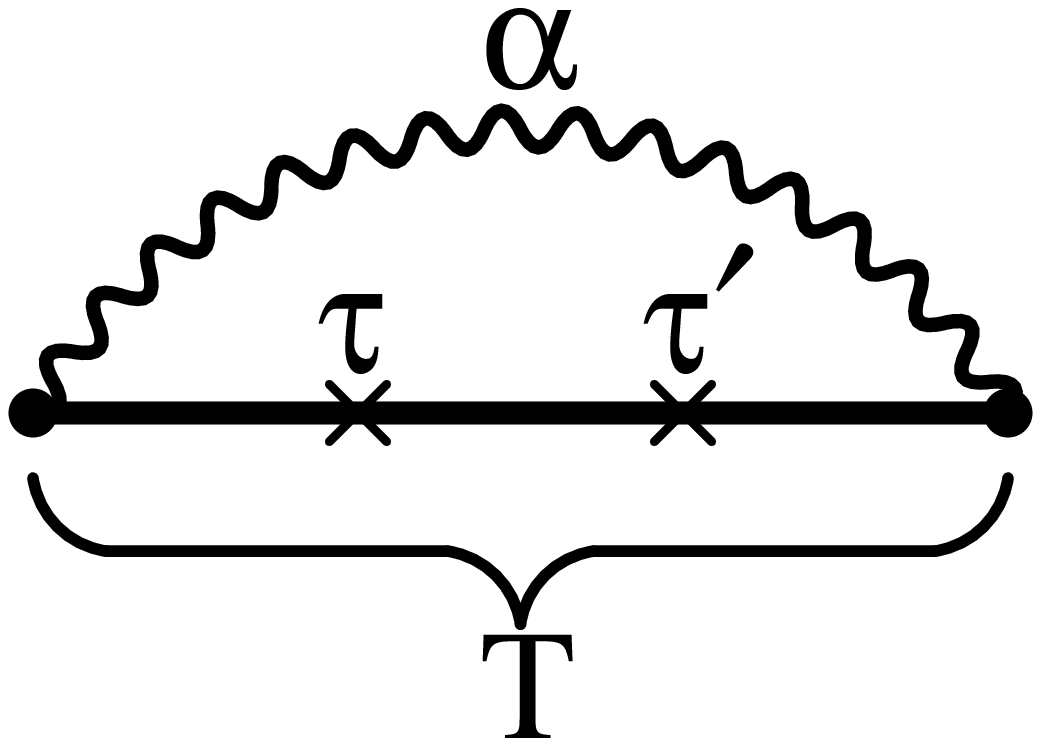,width=6cm}
\caption{Any set of diagrams with $\phi^4$-operator insertion 
can be obtained by
pinching a dummy photon propagator by setting the Feynman parameter
$\alpha \rightarrow 0$.}
\end{center}
\end{figure}
%
%ppppppppppppppppppppppppppppppppppppppppppppppppppppppppppppp
%ppppppppppppppppppppppppppppppppppppppppppppppppppppppppppppp
%
\begin{figure}
\begin{center}
\epsfile{file=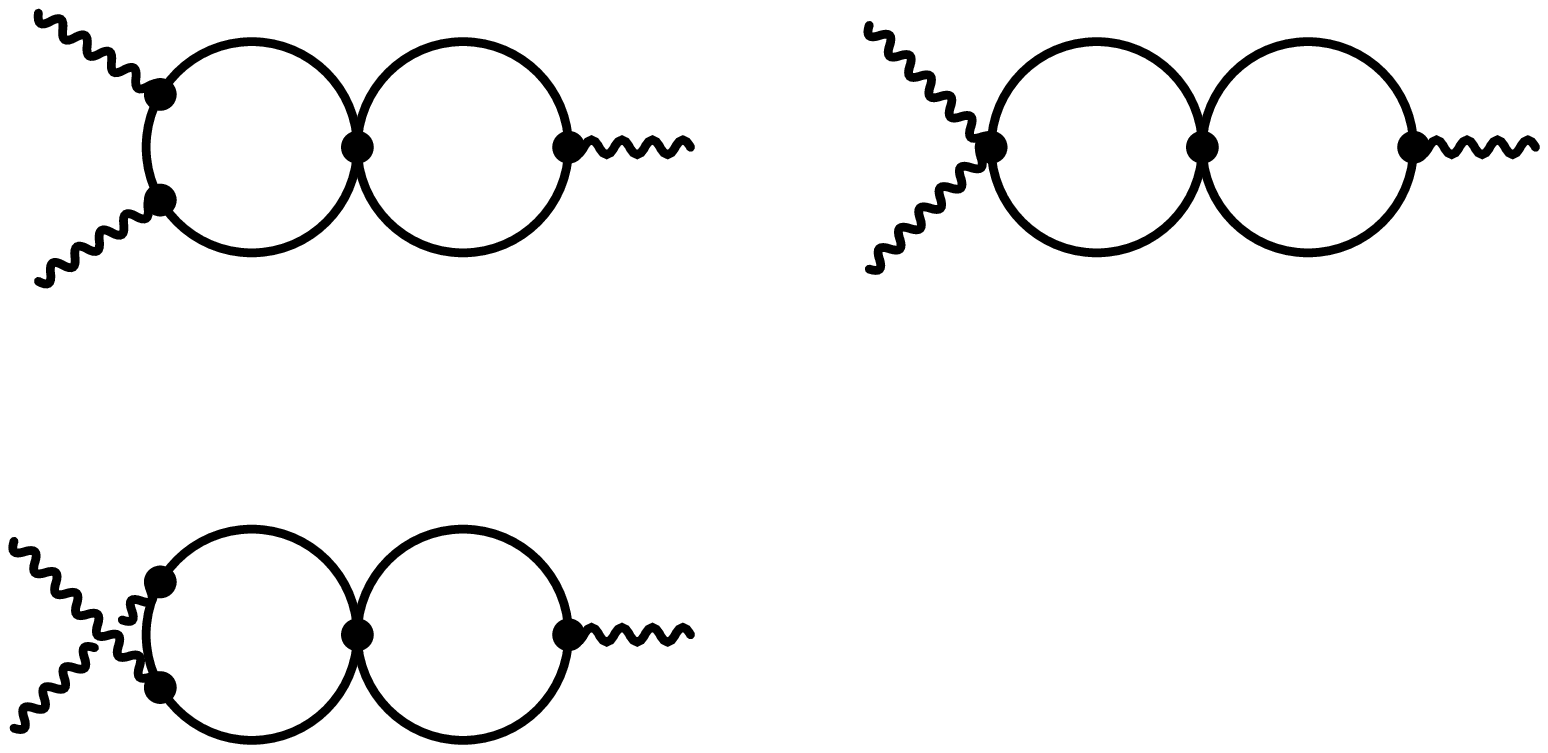,width=10cm}
\caption{A set of two-loop diagrams with a $\phi^4$-operator insertion.}
\end{center}
\end{figure}
%
%ppppppppppppppppppppppppppppppppppppppppppppppppppppppppppppp

The second example is the set of diagrams in Fig.16.
Also starting from eq.(\ref{opgenerfn}), we obtain
\begin{eqnarray}
G(k,\epsilon) &=&
(2\pi)^D \delta (\sum k_i ) \cdot i^2 
\left( \frac{1}{4\pi i} \right)^{D} \! \! \cdot
(i\lambda) (ie)^4 
\nonumber
\\
&& \times
\int^\infty_0 dT_1 
\int^\infty_0 dT_2
\, e^{-im^2(T_1+T_2)} 
\int^{T_1}_0 dt_1 dt_2 \int^{T_2}_0 dt_3 dt_4
\nonumber
\\
&& \times 
\Delta \, \exp \biggl[ \, \frac{1}{2}
\sum_{i \neq j} \biggl\{
- i k_i \cdot k_j G_B^{ij}
- 2 k_i \cdot \epsilon_j \partial_j G_B^{ij}
+ i \epsilon_i \cdot \epsilon_j \partial_i \partial_j G_B^{ij}
\biggl\}
\biggl] 
\end{eqnarray}
with $k_0 = p+p'$ and $\epsilon_0=0$.
This time the two-point function is obtained by sewing together two
scalar loops and pinching the photon propagator as in Fig.17.
Thus,
\begin{eqnarray}
G_B(\tau,\tau') &=& \left\{
\begin{array}{lcl}
|\tau-\tau'| - \frac{(\tau-\tau')^2}{T_1}
&~~&\tau,\tau' \in loop \, 1 \\
\rule{0mm}{6mm}
\tau - \frac{\tau^2}{T_1} + \tau' - \frac{\tau'^2}{T_2}
&~~&\tau \in loop \, 1, ~~ \tau' \in loop \, 2 \\
\rule{0mm}{6mm}
|\tau-\tau'| - \frac{(\tau-\tau')^2}{T_2}
&~~&\tau,\tau' \in loop \, 2
\end{array}
\right.
\end{eqnarray}
and
\begin{eqnarray}
\Delta = T_1 \, T_2 .
\end{eqnarray}
%ppppppppppppppppppppppppppppppppppppppppppppppppppppppppppppp
%
\begin{figure}
\begin{center}
\epsfile{file=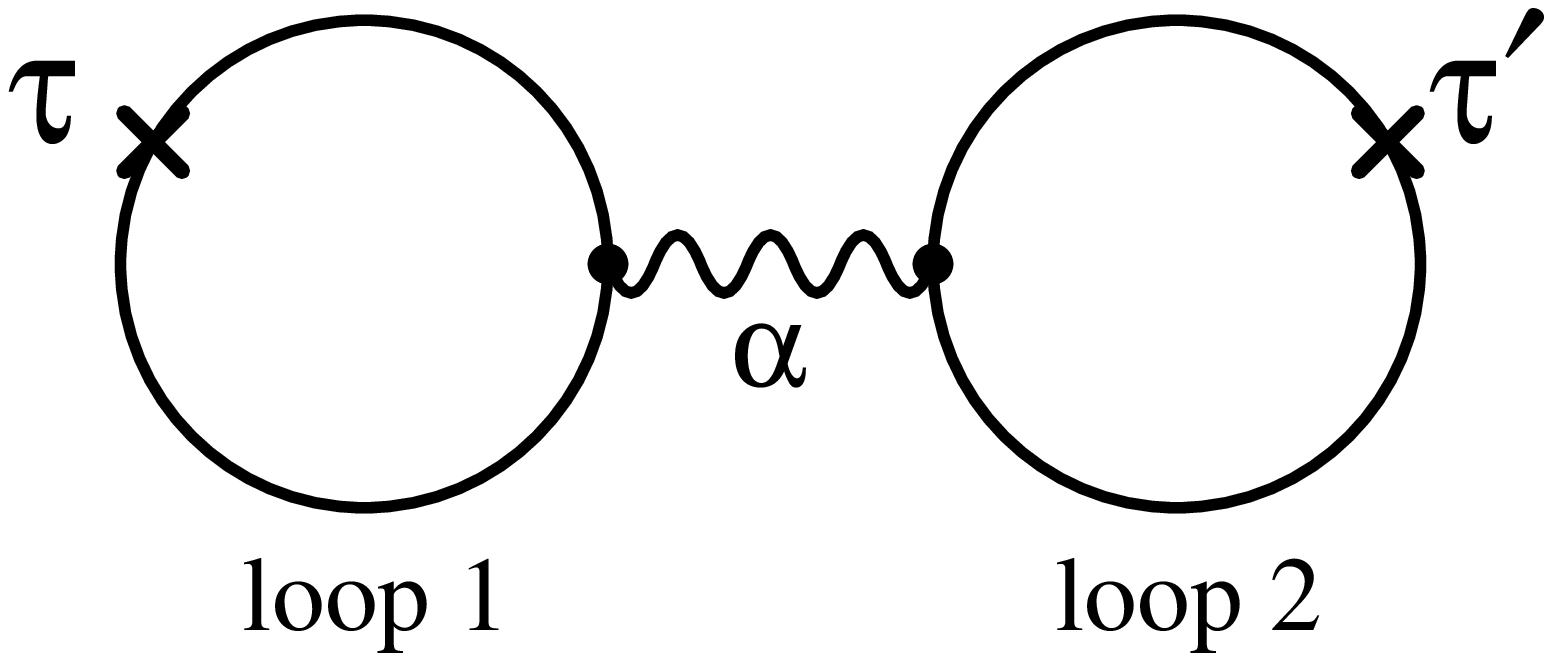,width=10cm}
\caption{The two-point function 
$G_B(\tau,\tau')$
of the diagrams in Fig.16 can be obtained by
sewing together two one-loop diagrams by a dummy photon propagator and
taking $\alpha \rightarrow 0$.}
\end{center}
\end{figure}
%
%ppppppppppppppppppppppppppppppppppppppppppppppppppppppppppppp

\section{Conclusion}
\cleqn

First of all, we conceive a set of diagrams connected by gauge
transformation as an entity expressed by a single path-integral.
The point is to assign proper time to the set of diagrams along the
charge flow and also express each photon propagator by Feynman parameter 
integral in coordinate space.
This enables one 
to find a general path-integral expression for any set
of diagrams starting from the quantum field theory.
At this stage, the resulting expression after integrating out $x(\tau)$
is equivalent to the Feynman parameter integral formula.
Simple rules for constructing the two-point function 
(correlation function on the worldline)
$G_B(\tau, \tau') \sim \left< x(\tau) \, x(\tau') \right>$
for a general set of diagrams is obtained.

Secondly, the path-integral expression allows us to use 
integration by parts technique both for
external and internal gauge vertices.
Manifestly gauge invariant form with respect to external photons can be
obtained before integrating over the proper time variables.
Surface terms can be neglected if the external scalars are on-shell.
The integration by parts technique reduces the number of independent
integrals, which can be interpreted as a non-trivial reshuffling of the 
original Feynman diagrams.

We have extended former trials to derive Bern-Kosower-type rule from
quantum field theory to
the general diagrams for scalar QED, 
in particular to the diagrams including external
scalar particles.
We have shown clear correspondence to the conventional Feynman rule,
which enabled us to avoid any ambiguity coming from the infinite
dimensionality of the path-integral approach.

The method for deriving the general
path-integral expression in section 2 can be 
straightforwardly extended to the case of spinor QED.

\section*{Acknowledgements}
One of the authors (Y.\ S.)\ is grateful to fruitful discussions with 
N.\ Ishibashi and M.\ Kitazawa.

\section*{Appendix A: Derivation of Eqs.(\ref{ialpha}) and (\ref{delta})}
\renewcommand{\theequation}{A.\arabic{equation}}
\cleqn

We show how to integrate over $x_i$'s in eq.(\ref{ialpha1}):
\begin{eqnarray}
I(\alpha ) \equiv \int [dx_i]
\, 
\exp \left[ - \frac{i}{4} \sum^n_{i,j=1} x_i \cdot x_j \, 
A_{ij}(\alpha ) + i \sum^n_{i=1} k_i \cdot x_i \right] .
\end{eqnarray}

First, insert the identity
\begin{eqnarray}
1 = \int d^Dc \, \, \delta \left( \sum^n_{i=1} x_i -c \right) ,
\end{eqnarray}
and shift all vertices as $x_i^\mu \rightarrow x_i^\mu + c^\mu /n$. 
We have
\begin{eqnarray}
I(\alpha ) &=& \int [dx_i]
\int d^Dc \, \, \delta \left( \sum x_i \right) \,
\exp \left[ - \frac{i}{4} \sum x_i \cdot x_j \, 
A_{ij} + i \sum k_i \cdot x_i 
+ \frac{i}{n} \sum k_i \cdot c
\right] 
\\
&=& 
(2\pi)^D \delta \left( \sum k_i \right) 
\cdot n^D
\int [dx_i]
\, \delta \left( \sum x_i \right)
\exp \left[ - \frac{i}{4} \sum x_i \cdot x_j \, 
A_{ij} + i \sum k_i \cdot x_i \right] 
\end{eqnarray}
We may further shift 
$x_i^\mu \rightarrow x_i^\mu - y^\mu /n$:
\begin{eqnarray}
I(\alpha) = 
(2\pi)^D \delta \left( \sum k_i \right) 
\cdot n^D
\int [dx_i]
\, \delta \left( \sum x_i - y \right) \,
\exp \left[ - \frac{i}{4} \sum x_i \cdot x_j \, 
A_{ij} + i \sum k_i \cdot x_i \right] .
\end{eqnarray}
It is independent of $y$.
Again insert
\begin{eqnarray}
1 = i \left( \frac{\beta}{4\pi i} \right)^{D/2}
\int d^Dy \, e^{-i\beta y^2 /4},
\end{eqnarray}
and integrate over $y$.
Thus,
\begin{eqnarray}
I(\alpha) &=& 
(2\pi)^D \delta \left( \sum k_i \right) 
\cdot i \left( \frac{\beta}{4\pi i} \right)^{D/2} n^D
\nonumber \\ 
&& ~~~ \times
\int [dx_i]
\, \exp \left[ - \frac{i}{4} \sum x_i \cdot x_j \, 
A'_{ij} + i \sum k_i \cdot x_i \right] ,
\end{eqnarray}
where $A'_{ij} = A_{ij} + \beta$.
Now the zero-mode is removed.
We may integrate over $x_i$'s, and noting the fact 
$\det A' = n\beta \cdot\det' A$, we obtain 
eqs.(\ref{ialpha}) and (\ref{delta}) with 
$Z = A'^{-1}$.
(It is necessary to transform
$Z_{ij}$ appropriately
for obtaining $Z$ in zero-diagonal level scheme; see Appendix B.)

\section*{Appendix B: Properties of $Z_{ab}$}
\renewcommand{\theequation}{B.\arabic{equation}}
\cleqn

{\large {\bf Definition}}
\\
\\
For a given scalar QED diagram without seagull vertex,
$Z_{ab}$ is defined by
\begin{eqnarray}
g^{\mu \nu} Z_{ab} \equiv \left( -\frac{i}{4} \right) ~
\frac{
\displaystyle
\int [d^Dx_c] \, (x_a-x_b)^\mu (x_a-x_b)^\nu \,
\exp \biggl[ -\frac{i}{4} \sum_{c,d}
x_c \cdot x_d A_{cd} \biggl]
}
{
\displaystyle
\int [d^Dx_c] \,
\exp \biggl[ -\frac{i}{4} \sum_{c,d}
x_c \cdot x_d A_{cd} \biggl]
}
.
\label{Bdefz}
\end{eqnarray}
On both sides of each vertex $i$ dummy vertices
$i'$ and $i''$ are inserted as shown in Fig.6.
Here,
$a,b,c,d$ denote vertices including dummy vertices ($i$, $i'$, and
$i''$).
The matrix $A$ represents the topology of the diagram, and
is defined by
\begin{eqnarray}
\sum_{c,d} x_c \cdot x_d \, A_{cd}
\equiv
\sum_{(cd)}
\frac{(x_c-x_d)^2}{\alpha_{(cd)}}, 
\end{eqnarray}
where $\alpha_{(cd)}$ denotes the Feynman parameter of the propagator
connecting the vertices $c$ and $d$.
\\
\\
{\large {\bf Methods for Calculating $Z_{ab}$}\footnote{
$Z_{ab}$ can also be computed using graph-theoretical 
formula\cite{lam}. 
}
}
\\
\\
In order to obtain $Z_{ab}$ from the matrix $A$, first
one may as well reduce the size of $A$ by eliminating
all external vertices in the diagram (but $a$
and/or $b$ if it is external) using the associativity property
(\ref{assoc}) of
propagator $K$.
Then, there are several ways to calculate $Z_{ab}$
from the reduced $A$. 
We exemplify two such methods here.
\\
\\
(Method 1)
Let $T$ be a matrix defined by
\begin{eqnarray}
T_{ab} = 1
~~~~~
\mbox{for}~\forall a,b,
\end{eqnarray}
and define $Z' \equiv (A+\beta T)^{-1}$. 
$Z'$ is well-defined as long as $\beta \neq 0$.
Then $Z_{ab}$ can be obtained using (\ref{transfz}) as
\begin{eqnarray}
Z_{ab} = Z'_{ab} - \frac{1}{2}(Z'_{aa}+Z'_{bb}).
\end{eqnarray}
Obviously the diagonal elements vanish.
$Z$ is independent of $\beta$ so one may 
simplify calculation by taking
$\beta \rightarrow \infty$ after getting $Z'$.
\\
\\
(Method 2)
Let $\tilde{A}$ be a submatrix of
$A$ obtained by deletion of the $c$-th row and $c$-th column.
One may choose any vertex $c$ for this purpose.
(This corresponds to fixing the coordinate of $c$ to be $x_c=0$
in eq.(\ref{Bdefz}).)
$\tilde{A}$ can be inverted, so define
\begin{eqnarray}
Z'_{ab} = \left\{ 
\begin{array}{cc}
(\tilde{A}^{-1})_{ab}&\mbox{for}~a,b \neq c \\
0&\mbox{otherwise}
\rule{0mm}{8mm}
\end{array}
\right. .
\end{eqnarray}
Then $Z_{ab}$ can be obtained as
\begin{eqnarray}
Z_{ab} = Z'_{ab} - \frac{1}{2}(Z'_{aa}+Z'_{bb}).
\end{eqnarray}
\\
\\
{\large {\bf Properties}}
\\
\begin{eqnarray}
&&
Z_{ab} = Z_{ba}
\\
&&
Z_{aa} = 0
\\
&&
\lim_{u'_i \rightarrow +0}
Z_{i'a} =
\lim_{u''_i \rightarrow +0}
Z_{i''a} =
Z_{ia}
\label{proof1}
\\
&&
\lim_{u'_i \rightarrow +0}
\frac{Z_{i'a}-Z_{ia}}{u'_i}
= \lim_{u'_i \rightarrow +0} 
\frac{\partial}{\partial u'_i} Z_{i'a}
\rule{0mm}{1.2cm}
\label{proof2}
\\
&&
\lim_{u''_i \rightarrow +0}
\frac{Z_{ia}-Z_{i''a}}{u''_i}
= - \lim_{u''_i \rightarrow +0}
\frac{\partial}{\partial u''_i} Z_{i''a}
\rule{0mm}{1.2cm}
\label{proof3}
\\
&&
\lim_{u'_i \rightarrow +0} \,
\frac{\partial}{\partial u'_i} Z_{i'i}
\rule{0mm}{1.2cm}
= \lim_{u''_i \rightarrow +0} \,
\frac{\partial}{\partial u''_i} Z_{i''i}
\rule{0mm}{1.2cm}
= - \frac{1}{2}
\label{proof4}
\\
&&
\lim_{\begin{array}{c} 
         \scriptstyle u'_i \rightarrow +0\\
         \scriptstyle u'_j \rightarrow +0
      \end{array}}
\frac{Z_{i'j'}-Z_{ij'}-Z_{i'j}+Z_{ij}}{u'_i u'_j} 
= \left\{ 
\begin{array}{cc}\displaystyle
\lim_{\begin{array}{c} 
         \scriptstyle u'_i \rightarrow +0\\
         \scriptstyle u'_j \rightarrow +0
      \end{array}}
\frac{\partial^2}{\partial u'_i \partial u'_j}
Z_{i'j'}
&~~~\mbox{for}~i \neq j
\\
\infty 
\rule{0mm}{8mm}
&~~~\mbox{for}~i=j
\end{array}
\right.
\label{end}
\end{eqnarray}
\\
\\
{\large (Proof)} \\
Eq.(\ref{proof1}): Use eq.(\ref{propk2}). \\
Eqs.(\ref{proof2}) and (\ref{proof3}): Use eq.(\ref{proof1}). \\
Eq.(\ref{proof4}): 
\begin{eqnarray}
\lim_{u'_i \rightarrow +0} \,
\frac{\partial}{\partial u'_i} Z_{i'i}
\rule{0mm}{1.2cm}
= \lim_{u'_i \rightarrow +0}
\frac{Z_{i'i}-Z_{ii}}{u'_i}
= \lim_{u'_i \rightarrow +0}
\frac{Z_{i'i}}{u'_i} .
\end{eqnarray}
Then substituting the definition (\ref{Bdefz}), the integrand will be
\begin{eqnarray}
&&
\frac{(x'_i-x_i)^\mu (x'_{i}-x_i)^\nu}{u'_i} \, 
\exp \biggl[ -\frac{i}{4} \frac{(x'_i-x_i)^2}{u'_i} \biggl]
=
(x'_i-x_i)^\mu \, 2i \, \frac{\partial}{\partial x'_{i \, \nu}} \, 
\exp \biggl[ -\frac{i}{4} \frac{(x'_i-x_i)^2}{u'_i} \biggl]
\nonumber
\\
&&=
- 2i \, g^{\mu\nu} \,
\exp \biggl[ -\frac{i}{4} \frac{(x'_i-x_i)^2}{u'_i} \biggl] ,
\end{eqnarray}
where in the last line we integrated by parts with respect to $x'^\nu_i$.
Thus, the numerator will be proportional to the donominator in (\ref{Bdefz}).

\section*{Appendix C: Sample Calculation}
\renewcommand{\theequation}{C.\arabic{equation}}
\cleqn

%ppppppppppppppppppppppppppppppppppppppppppppppppppppppppppppp
%
\begin{figure}
\begin{center}
\epsfile{file=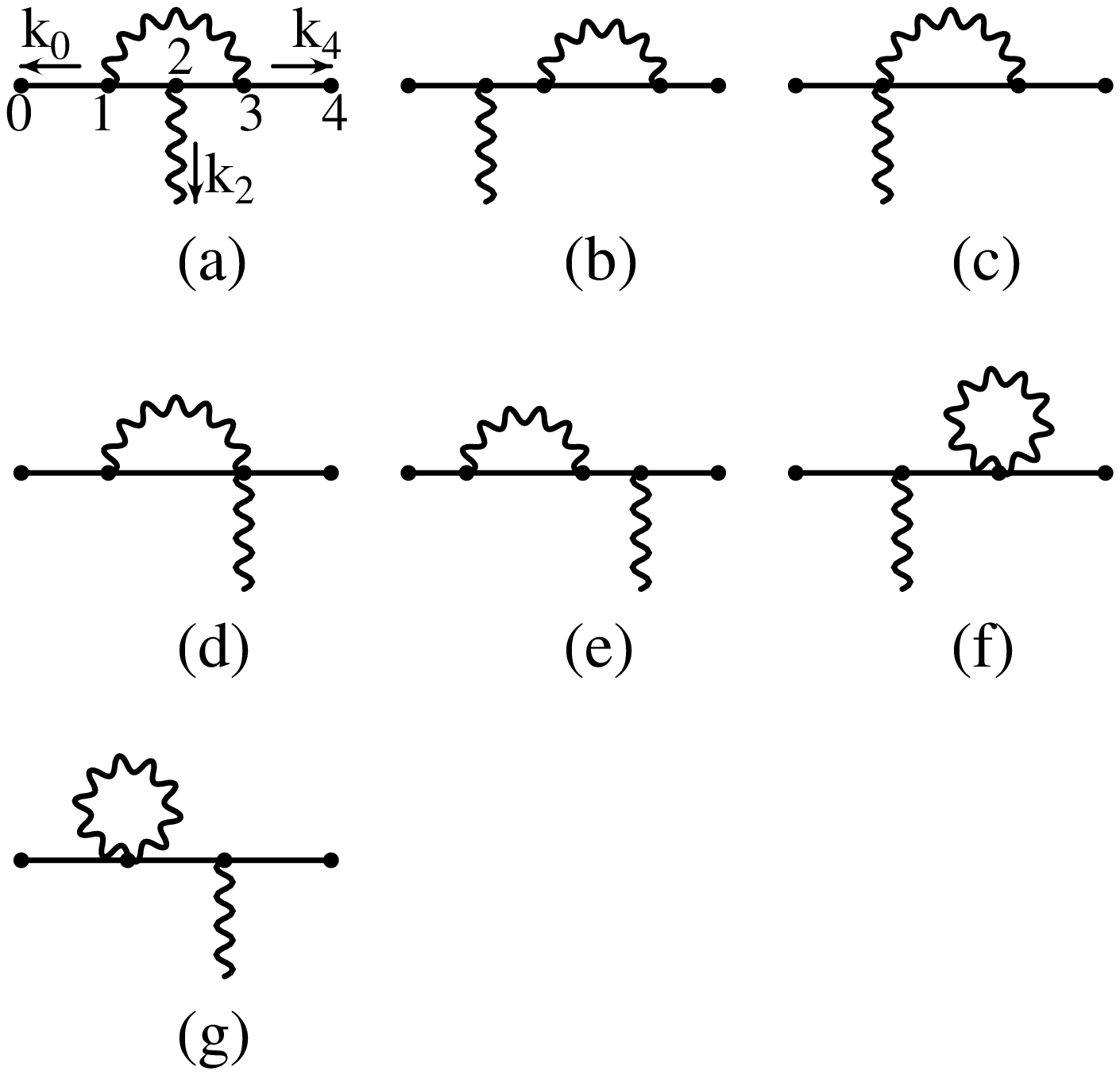,width=10cm}
\caption{The set of diagrams calculated in Appendix C.}
\end{center}
\end{figure}
%
%ppppppppppppppppppppppppppppppppppppppppppppppppppppppppppppp

In this appendix we apply the Bern-Kosower-type rule to calculation
of the set of diagrams shown in Fig.18.
According to eq.(\ref{rule1}), the Green function is given by
\begin{eqnarray}
G_S(k_0,k_2,k_4,\epsilon_2) &=&
(2\pi)^D \delta
\biggl( \sum_{i=0}^4 k_i \biggl) 
\cdot i \left( \frac{1}{4\pi i}
\right)^{D/2} 
\frac{1}{2} \, (ie)^3 
\nonumber \\
&& \times
\int^\infty_0 d\alpha \, e^{-i(\lambda^2-i0)\alpha} 
\int^\infty_0 dT \, e^{-i(m^2-i0)T} 
\int^{T}_0 dt_1 dt_2 dt_3 
\, {\cal K}_{red},
\end{eqnarray}
where $\lambda$ is the photon mass.
${\cal K}_{red}$ is obtained from $\cal K$ in eq.(\ref{gkf}) after the
manipulation 1)-5):
\begin{eqnarray}
{\cal K}_{red} &=& \Delta^{-D/2} \,
\biggl[
-\sum_{i=0}^4 k_i^\mu \partial_1 G_B^{i1} 
\sum_{j=0}^4 k_{j\mu} \partial_3 G_B^{j3} 
\sum_{l=0}^4 \epsilon_2' \cdot k_l \partial_2 G_B^{l2}
+ i \partial_1 \partial_2 G_B^{12}  \sum_{j=0}^4 \epsilon_2' \cdot 
k_j \partial_3 G_B^{j3}
\nonumber
\\ &&
+ i \partial_2 \partial_3 G_B^{23} \sum_{j=0}^4 \epsilon_2' \cdot 
k_i \partial_1 G_B^{i1}
+ i D \partial_1 \partial_3 G_B^{13} \sum_{l=0}^4 \epsilon_2' \cdot 
k_l \partial_2 G_B^{l2} 
\biggl]
\, \exp \biggl[ -\frac{i}{2} \sum_{i \neq j} k_i \cdot k_j G_B^{ij}
\biggl] .
\end{eqnarray}
Here, we choose
\begin{eqnarray}
\epsilon_2'^\mu = \epsilon_2^\mu 
- \frac{\epsilon_2 \cdot k_2}{k_2^2} k_2^\mu ,
\end{eqnarray}
so that $\epsilon_2' \cdot k_2 = 0$.

Now we integrate by parts with respect to $t_2$:
\begin{eqnarray}
{\cal K}_{red} \rightarrow & \Delta^{-D/2}
\biggl[ &
( k_0 \partial_1 G_B^{01} + k_2 \partial_1 G_B^{21} + 
k_4 \partial_1 G_B^{41} )
\cdot
( k_0 \partial_3 G_B^{03} + k_2 \partial_3 G_B^{23} + 
k_4 \partial_3 G_B^{43} ) 
\nonumber
\\ &&
\times \,
\epsilon_2' \cdot ( k_0 \partial_2 G_B^{02} + k_4 \partial_2 G_B^{42} )
\nonumber
\\ &&
- \partial_1 G_B^{12} \,
\epsilon_2' \cdot ( k_0 \partial_3 G_B^{03} + k_4 \partial_3 G_B^{43} )
\, k_2 \cdot ( k_0 \partial_2 G_B^{02} + k_4 \partial_2 G_B^{42} )
\nonumber
\\&&
- \partial_3 G_B^{23} \,
\epsilon_2' \cdot ( k_0 \partial_1 G_B^{01} + k_4 \partial_1 G_B^{41} )
\, k_2 \cdot ( k_0 \partial_2 G_B^{02} + k_4 \partial_2 G_B^{42} )
\nonumber
\\&&
+ iD \, \partial_1 \partial_3 G_B^{13}
\epsilon_2' \cdot ( k_0 \partial_2 G_B^{02} + k_4 \partial_2 G_B^{42} )
\biggl]
\nonumber
\\&& ~~~
\times \exp [
-i ( k_0 \cdot k_2 G_B^{02} + k_0 \cdot k_4 G_B^{04} + k_2 \cdot k_4
G_B^{24}) ]
\end{eqnarray}
We do not integrate by parts with respect to $t_1$ or $t_3$; compare the 
discussion in subsection 4.c.
The delta function part in $\partial_1 \partial_3 G_B$ corresponds to
the tadpole diagrams (Fig.18(f)(g)).

Then we substitute the explicit forms of $\Delta$, $G_B^{ij}$, and their
derivatives: 
\begin{eqnarray}
\Delta &=& \alpha + |t_3-t_1|,
\\
G_B^{ij} &=& |t_i-t_j| - 
\frac{
[ |t_i-t_1| - |t_i-t_3| - |t_j-t_1| + |t_j-t_3| ]^2
}{4\Delta} ,
\\
\partial_j G_B^{ij} &=& - \mbox{sign} (t_i-t_j) +
\frac{1}{2\Delta}
[ |t_i-t_1| - |t_i-t_3| - |t_j-t_1| + |t_j-t_3| ] 
\nonumber
\\ && ~~~~~~~~~~ ~~~~~~~~~~ ~~~~~~~~~~\times 
[ \mbox{sign} (t_j-t_1) - \mbox{sign} (t_j -t_3) ] ,
\label{cdelgbj}
\\
\partial_1 \partial_3 G_B^{13} &=&
-2\delta (t_1-t_3) + \frac{1}{2\Delta} ,
\end{eqnarray}
where $t_0=0$ and $t_4=T$.
It is understood that $\mbox{sign}(0)=0$ in eq.(\ref{cdelgbj}).
Once the time ordering of $t_1$, $t_2$, and $t_3$ is fixed, we can
transform the integral variables using eq.(\ref{ivtransf1}).
The rest is same as the usual Feynman parameter integral.
We obtain, for example,
\begin{eqnarray}
G_S(t_1<t_2<t_3) &=&
(2\pi)^D \delta
( \sum k_i ) 
\cdot i \left( \frac{1}{4\pi i}
\right)^{D/2} \, (ie)^3 
\biggl[ \frac{i}{k_0^2-m^2} \, \frac{i}{k_4^2-m^2} \biggl]
\nonumber
\\ && \times
i \epsilon_2' \cdot (k_4-k_0)
\, \biggl[
(1-\omega ) \, I_1 + \omega \, I_2 +
(-i)^{-D/2} \, \Gamma (2-{\textstyle \frac{D}{2}}) \, I_3
\biggl]
\end{eqnarray}
where $\omega = - k_0 \cdot k_4 /m^2 > 1$, and
\begin{eqnarray}
I_1 &=& \int^1_0 dx \int^{1-x}_0 dy \, 2 (1-2x)(y^2-2y)
\times [x^2 + y^2 + 2\omega x y ]^{-1}
\nonumber
\\
&=& \frac{7}{6} \frac{1}{\omega -1} - 
\frac{1}{\sqrt{\omega^2-1}}
\biggl( \frac{3}{2} + \frac{7}{6} \frac{1}{\omega - 1} 
\biggl)
\mbox{arccosh} \, \omega
\\
I_2 &=& 
\int^1_0 dx \int^{1-x}_0 dy \, 
(1-x-y)(x+y-2)^2 \times
\biggl[ 
x^2+y^2+2\omega xy + \frac{\lambda^2}{m^2} (1-x-y) 
\biggl]^{-1}
\nonumber
\\
&=&
- \frac{1}{\sqrt{\omega^2-1}} 
\biggl[ \frac{35}{6} + 2 \log \frac{\lambda^2}{m^2} \biggl]
\mbox{arccosh} \, \omega
+ \frac{8}{\sqrt{\omega^2-1}}
\int^{\frac{1}{2}\mbox{arccosh} \, \omega}_0 
d\varphi \, \varphi \, \tanh \varphi
\\
I_3 &=&
\int^1_0 dx \int^{1-x}_0 dy \, 
\frac{1}{2} (1-x-y)
\times [m^2 (x^2 + y^2 + 2\omega xy) ]^{D/2-2}
\nonumber
\\ &=&
\frac{1}{12} + \frac{D-4}{4}
\biggl( -\frac{11}{18} + \frac{1}{6} \log \frac{m^2}{\mu^2}
- \frac{1}{6} \sqrt{ \frac{\omega+1}{\omega-1} }
\mbox{arccosh} \, \omega \biggl) .
\end{eqnarray}
We set the external scalars on-shell $k_0^2=k_4^2=m^2$ except for the
propagator factors in the above expressions.
$G_S$ for other time orderings can be calculated similarly.
(See below.)

Finally, if we are interested in the vertex function, we should
amputate the external scalars in the above example.
For this purpose, one should add the counter term for the wave
function correction first, which needs to be calculated separately.
After adding the counter term and amputating the external propagators, 
we find the vertex function at one-loop
(for on-shell external scalars) to be
\begin{eqnarray}
\epsilon_2^\mu \, \Gamma_\mu^{\mbox{\scriptsize 1-loop}} (k_0,k_4)
&=&
- \frac{e^2}{16\pi^2} \, \epsilon_2 \cdot (k_4-k_0) \,
\biggl[
\frac{9}{2(4-D)} 
- \frac{9}{4} ( \log \frac{m^2}{4\pi \mu^2} + \gamma_E ) 
+ \frac{19}{4} 
\nonumber
\\
&& + \frac{1}{\sqrt{\omega^2-1}}
\biggl( \frac{19}{12} - \frac{17}{4} \omega 
- 2\omega \log \frac{\lambda^2}{m^2} \biggl) 
\mbox{arccosh} \, \omega
\nonumber
\\ &&
+ \frac{8 \omega}{\sqrt{\omega^2-1}}
\int^{\frac{1}{2} \mbox{arccosh} \omega}_0
d\varphi \, \varphi \, \tanh \varphi
\biggl] .
\end{eqnarray}

%%%%%%% References

\end{document}